%% file: o3imbh.tex
\documentclass[onecolumn]{aa}
\bibpunct{(}{)}{;}{a}{}{,}
\usepackage{graphicx}
\usepackage{txfonts}
\usepackage{color}
\usepackage{calc}
\usepackage{amsmath,amssymb,graphicx}
\usepackage{tensor}
\usepackage{bm}
\usepackage{times}
\usepackage[colorlinks, linkcolor=blue, pdfborder={0 0 0}, breaklinks=true]{hyperref}
\usepackage{float}
\usepackage{dcolumn}
\usepackage[nolist,nohyperlinks]{acronym}
\usepackage{xspace}
\usepackage[abs]{overpic}
\usepackage{pict2e}
\usepackage{enumitem}
\usepackage[usenames,dvipsnames]{xcolor}
\usepackage[utf8]{inputenc}
\usepackage{gensymb}
\usepackage[capitalize]{cleveref}
\Crefname{figure}{Fig.}{Figs.}% {<type>}{<singular>}{<plural>}
\usepackage[normalem]{ulem}
\usepackage{array}
\usepackage{multirow}
\usepackage{subfigure}
\usepackage{tabularx}
\usepackage{etoolbox}
\usepackage{siunitx}
\usepackage{lineno}
\usepackage{booktabs}

\usepackage{scrextend}
\usepackage{textgreek}
\newcommand{\Msun}{\ensuremath{\mathrm{M}_\odot}}

\newcommand{\christmasevent}{191225\_215715}
\newcommand{\valentinesdaynoise}{200214\_224526}
\newcommand{\januaryevent}{200114\_020818}
\newcommand{\decembertwentythirdnoise}{191223\_014159}
\newcommand{\septembernoise}{190924\_232654}

\begin{document}

   \title{Search for intermediate mass black hole binaries in the third observing run of Advanced LIGO and Advanced Virgo}

   \author{The LVK Collaboration (full author list in Appendix C)}

%\date[\relax]{Compiled: \today}

\abstract{
Intermediate-mass black holes (IMBHs) span the approximate mass range 
$100$--$10^5$\,\Msun, between black holes (BHs) formed by stellar collapse
and the supermassive BHs at the centers of galaxies.  Mergers of IMBH binaries are 
the most energetic gravitational-wave sources accessible by the terrestrial detector network. 
Searches of the first two observing runs of Advanced LIGO and Advanced Virgo did not 
yield any significant IMBH binary signals.  
In the third observing run (O3), the increased network sensitivity enabled the detection 
of GW190521, a signal consistent with a binary merger of mass $\sim 150$\,\Msun\, providing
direct evidence of IMBH formation. 
Here we report on a 
dedicated search of O3 data for further IMBH binary mergers, combining both modelled 
(matched filter) and model independent search methods. 
We find some marginal candidates, but none are sufficiently significant to indicate 
detection of further IMBH mergers. 
We quantify the sensitivity of the individual search methods and of the combined search
using a suite of IMBH binary signals obtained via numerical relativity, 
including the effects of spins misaligned with the binary orbital axis, and present 
the resulting upper limits on astrophysical merger rates.  Our most stringent limit is for 
equal mass and aligned spin BH binary of total mass $200 \Msun$ and effective aligned spin 0.8 
at {$0.056$\,\si{Gpc^{-3}yr^{-1}}} (90\protect\% confidence), a factor of 3.5
more constraining than previous LIGO-Virgo limits.  We also update the estimated rate of
mergers similar to GW190521 to {$0.08$\,\si{Gpc^{-3}yr^{-1}}}.
}

% \abstract{}{}{}{}{}
% 5 {} token are mandatory

  % aims heading (mandatory)

  % conclusions heading (optional), leave it empty if necessary

%\titlerunning{O3 IMBH search}
%\authorrunning{LVK}

\maketitle

% ======================
%  ACRONYMS
% ======================

\acrodef{LIGO}[LIGO]{Laser Interferometer Gravitational-Wave Observatory}
\acrodef{LSC}[LSC]{LIGO Scientific Collaboration}
\acrodef{aLIGO}{Advanced Laser Interferometer Gravitational wave Observatory}
\acrodef{aVirgo}{Advanced Virgo}
\acrodef{LVC}[LIGO-Virgo]{LIGO and Virgo Collaborations}
\acrodef{LHO}[LHO]{LIGO-Hanford}
\acrodef{LLO}[LLO]{LIGO-Livingston}
\acrodef{O3}[O3]{third observing run}
\acrodef{O2}[O2]{second observing run}
\acrodef{O1}[O1]{first observing run}

\acrodef{GW}[GW]{gravitational wave}
\acrodef{CBC}[CBC]{compact binary coalescence}
\acrodef{BH}[BH]{black hole}
\acrodef{NS}[NS]{neutron star}
\acrodef{BBH}[BBH]{binary black hole}
\acrodef{BNS}[BNS]{binary neutron star}
\acrodef{NSBH}[NSBH]{neutron star--black hole binary}
\acrodef{IMBH}[IMBH]{intermediate mass black hole}
\acrodef{IMBHB}[IMBHB]{intermediate mass black hole binary}
\acrodef{NR}[NR]{numerical relativity}
\acrodef{PN}[PN]{post-Newtonian}
\acrodef{CWB}[cWB]{coherent WaveBurst}
\acrodef{SNR}[SNR]{signal-to-noise ratio}
\acrodef{FAR}[FAR]{false alarm rate}
\acrodef{IFAR}[IFAR]{inverse false alarm rate}
\acrodef{FAP}[FAP]{false alarm probability}
\acrodef{PSD}[PSD]{power spectral density}

\newcommand{\PN}[0]{\ac{PN}\xspace}
\newcommand{\BBH}[0]{\ac{BBH}\xspace}
\newcommand{\BNS}[0]{\ac{BNS}\xspace}
\newcommand{\BH}[0]{\ac{BH}\xspace}
\newcommand{\NR}[0]{\ac{NR}\xspace}
\newcommand{\SNR}[0]{\ac{SNR}\xspace}
\newcommand{\aLIGO}[0]{\ac{aLIGO}\xspace}
\newcommand{\PE}[0]{\ac{PE}\xspace}
\newcommand{\IMR}[0]{\ac{IMR}\xspace}
\newcommand{\PDF}[0]{\ac{PDF}\xspace}
\newcommand{\GR}[0]{\ac{GR}\xspace}
\newcommand{\PSD}[0]{\ac{PSD}\xspace}
%
%________________________________________________________________
\input{introduction.tex}

%__________________________________________________________________
\input{data.tex}
\input{searches.tex}
\input{results.tex}
\input{rate.tex}
\input{conclusion.tex}
\begin{appendix}
\input{appendix.tex}
\input{s200114f.tex}

\input{LSC-Virgo-KAGRA-Authors-Feb-2021.tex}
\end{appendix}
\begin{acknowledgements}
\input{P2000488_v5.tex}

\end{acknowledgements}
\clearpage
\bibliographystyle{aa}
\bibliography{reference.bib}

%-------------------------------------------------------------------

\end{document}

%% file: introduction.tex
\section{Introduction}
\label{sec:intro}

Black holes are classified according to their masses: stellar-mass \acp{BH} are those with mass 
below $\sim 100\,\Msun$, formed by stellar collapse, while supermassive \acp{BH} \citep{Ferrarese_2005} at the centers of galaxies
have masses above $10^5\,\Msun$. Between stellar-mass and supermassive \acp{BH} is the realm of \acp{IMBH} -- \acp{BH} with 
masses in the range $100-10^5\,\Msun$ \citep{2004cbhg.symp...37V,2004IJMPD..13....1M,Ebisuzaki:2001qm, 2017mbhe.confE..51K, 2020ARA&A..58...27I}. 

Stellar evolution models suggest that \acp{BH} with mass up to $\sim 65\,\Msun$ are the result of 
core-collapse of massive stars \citep{2017ApJ...836..244W, 2018MNRAS.474.2959G, 2019ApJ...878...49W, 2019ApJ...887...53F, 2020ApJ...888...76M, 2020ApJ...902L..36F}. The final fate of the star is determined by the mass of the helium core alone. Stars with helium core mass in the range $\sim 32-64\,\Msun$ undergo pulsational pair-instability leaving behind remnant BHs of mass below $\sim 65\,\Msun$ \citep{1964ApJS....9..201F, 1967PhRvL..18..379B}. When the helium core mass is in the range $\sim 64-135\,\Msun$, pair-instability drives the supernova explosion and leaves no remnant; 
while stars with helium core mass greater than $\sim 135\,\Msun$ are expected to directly collapse to 
intermediate-mass \acp{BH}. Thus, 
pair-instability (PI) prevents the formation of heavier \acp{BH} from core-collapse, and suggests a mass gap
between $\sim 65-120\,\Msun$ in the BH population known as PI supernova (PISN) mass gap ~\citep{1984ApJ...280..825B, 2007Natur.450..390W,woosley2021pairinstability}. 
Possible \ac{IMBH} formation channels also include the direct collapse of massive first-generation, low-metallicity 
Population III stars \citep{2001ApJ...550..372F, 2003ApJ...591..288H, 2017MNRAS.470.4739S,2001ApJ...551L..27M, 2002ApJ...567..532H}, and multiple, hierarchical 
collisions of stars in dense young star clusters \citep{2002MNRAS.330..232C, 2006ApJ...637..937O, 2015MNRAS.454.3150G,2016MNRAS.459.3432M}, among others. It is not currently known how supermassive black holes form. Hierarchical merger of IMBH systems in a 
dense environment is among the putative formation channels for supermassive BHs \citep{King:2004ri, 2010A&ARv..18..279V, Merzcua:2017, 2017mbhe.confE..51K}. 

Several \ac{IMBH} candidates are suggested by electromagnetic observations, but lack conclusive confirmation 
\citep{Greene:2019vlv}. Observations include direct kinematical measurement of the mass of the 
central BH in massive star clusters and galaxies \citep{Merzcua:2017, 2002MNRAS.330..232C,AtakanGurkan:2003hm, Anderson_2010, Baumgardt:2002rc, Pasham:2015tca, Vitral2021}.
Other possible evidence for \ac{IMBH} includes extrapolation of scaling relations between the masses of host galaxies and their central supermassive BH 
to the mass range of globular clusters \citep{Graham:2012bs,2013ApJ...764..151G, Kormendy:2013dxa}.
In addition, observations of characteristic imprints on the surface brightness, mass-to-light ratio and/or line-of-sight velocities also suggest 
that dense globular clusters harbour IMBHs \citep{vandenBosch:2005uk, Gebhardt_2005, Noyola_2008, L_tzgendorf_2011, K_z_ltan_2017}. 
Controversy exists regarding the interpretation of these observations, as some of them can also be explained by a high concentration of 
stellar-mass BHs or the presence of binaries \citep{Baumgardt:2002rc, Anderson_2010, Lanzoni_2013}.
Empirical mass scaling relations of quasi-periodic oscillations in luminous X-ray sources have also provided evidence for IMBHs \citep{Remillard:2006}. 
Ultraluminous X-ray sources exceed the Eddington luminosity of an accreting stellar-mass BH \citep{Kaaret:2017tcn, Farrell:2010bf}.  An accreting IMBH is a favored explanation in several cases \citep{2001MNRAS.321L..29K, 2004IJMPD..13....1M}. 
However, neutron stars or stellar-mass black holes emitting above their Eddington luminosity could also account for such observations \citep{2014Natur.514..202B, 2017Sci...355..817I}. The strongest IMBH candidate amongst them is HLX-1, an hyper-luminous X-ray source indicating an IMBH mass of ${\sim} 0.3-30\times 10^4\, M_\odot$  \citep{Farrell:2010bf, 2009ApJ...705L.109G, 2011ApJ...743....6S,2012Sci...337..554W, 2015MNRAS.446.3268C,2012MNRAS.420.3599S}. 
In~\citet{Lin_2018}, an intermediate-mass black hole candidate was found in a tidal disruption event in a massive star cluster. More recently, in \citet{Paynter2021}, there was a claim of an IMBH detection through a gravitationally lensed gamma ray burst. 

The Advanced LIGO \citep{TheLIGOScientific:2014jea} and Advanced Virgo \citep{TheVirgo:2014hva} interferometric \ac{GW} detectors have completed three observing runs between September 2015 and March 2020. 
The third observing run of Advanced LIGO and Advanced Virgo, O3, extended from April 1st, 2019, 15:00 UTC to 
March 27th, 2020 17:00 UTC.  The recently released second gravitational-wave transient catalog %GWTC-2 
provided a comprehensive summary of significant compact binary coalescence events observed up to 
October 1st, 2019~\citep{GWTC2}, reporting a total of 50 events.  The corresponding binary black hole (BBH) population analysis of 
\citet{o3apop} indicates that $99\%$ of primary \ac{BH} masses lie below $m_{99\%} \sim 60\,\Msun$:
thus, the large majority of merging \ac{BH} have masses below a limit of $\sim 65\,\Msun$ consistent with 
expectations from PI. 

Near the beginning of O3, the first intermediate-mass black hole coalescence event, GW190521~\citep{GW190521Adiscovery}, was 
observed. This GW signal was consistent with 
a coalescence of black holes of $85^{+21}_{-14} \Msun$, and $66^{+17}_{-18} \Msun$ 
which resulted 
in a remnant black hole of $142^{+28}_{-16} \Msun$ falling in the mass range of 
intermediate-mass black holes. GW190521 provided the first conclusive evidence for the formation of an
IMBH below $10^3\,\Msun$. It is a massive binary black hole system with an IMBH remnant and 
a primary BH in the PISN mass gap with high confidence \citep{GW190521Aastro} although, see \citet{Fishbach_2020, Nitz:2020mga} for an alternative interpretation.
The discovery triggered a variety of investigations regarding the evolution models and the subsequent 
mass gap in the BH population.  It also suggested a possibility of the formation of massive 
BHs (>$100\,\Msun$) via hierarchical merger scenario in a dense environment \citep{GW190521Aastro, kimball2020evidence}. 
 
The Advanced LIGO - Advanced Virgo detectors are
sensitive to the lower end of the IMBH binary mass range, potentially making IMBHs detectable out to cosmological distances, as is evident from GW190521. Observation of IMBH binary systems are not only interesting for massive BH formation channels, but they act as a 
perfect laboratory to test general relativity \citep{TheLIGOScientific:2016src,Yunes:2016jcc,Yunes:2013dva,Gair:2012nm}. 
Massive BH coalescences produce louder mergers and ringdown signals in the sensitive band of the advanced GW detectors.
Furthermore, these can display prominent higher-order modes that confer GWs a more complex morphology that
can significantly deviate from a canonical chirp \citep{Calderon_Bustillo_2020}. Observations of higher-order modes help
to test general relativity and fundamental properties of BHs such as the no-hair theorem~\citep{Kamaretsos:2011um,Meidam:2014jpa,Thrane:2017lqn,Carullo:2018sfu} and BH kick measurements~\citep{Gonzalez:2006md,Campanelli:2007cga,CalderonBustillo:2018zuq}. These IMBHs might be multi-band events observable by both LIGO/Virgo and {\em LISA} \citep{2017arXiv170200786A}, and could provide novel probes of cosmology and contribute to the stochastic background~\citep{Fregeau_2006, Miller_2009, Jani2020, 2021ApJ...909L..23E}.

The GW signal from a massive BBH coalescence is evident as a short-duration waveform with little inspiral and mostly merger-ringdown signal, falling in the low-frequency region of the advanced detectors.  
With initial GW detectors \citep{Virgo:2012aa, Aasi:2014bqj}, the IMBH binary searches were restricted to probe the merger-ringdown phase of the coalescing BBH system, using the model waveform independent coherent WaveBurst (cWB) \citep{Klimenko:2004qh, Klimenko:2005xv, Klimenko:2006rh}, and a ringdown templated search \citep{Aasi:2014bqj}. Improvement in the detector sensitivity at low frequencies in the advanced era made IMBH binaries a target for a matched filtering search that would probe the short inspiral phase. In~\citet{Abbott:2017iws}, we used a combined search with the matched filtering GstLAL~\citep{gstlalmethods1, Chaddtdphi, sachdev2019gstlal} search and model independent cWB \citep{Klimenko:2011hz, Klimenko:2015ypf}. This combined search was further extended with an additional matched filtering PyCBC search~\citep{Usman:2015kfa,Allen:2004gu,Canton:2014ena,Nitz:2017svb} in \citet{Salemi:2019ovz}
using the data from the first two observing runs.  No significant IMBH binary event was found in these searches.

While all the previous matched filtering searches were generic BBH searches, the improvements in the detector sensitivity at low frequencies and the IMBH merger signals' short duration nature motivated us to use matched filter searches targeted to the IMBH mass-spin parameter space. Here, we carry out an IMBH binary search using the entire year-long third observing run, O3, of the Advanced LIGO and Advanced Virgo detector network with a combined search using three search algorithms: two matched-filtering based focused IMBH binary searches, using the PyCBC and GstLAL libraries, and the minimally modelled time-frequency based cWB search. We 
search for massive binary systems with at least one component above the expected PISN mass gap limit of $65\,\Msun$, and with an IMBH remnant.
GW190521 remains as the most significant candidate in the combined search; no other event is comparably significant. We provide the results from the combined search with the next most significant events and follow up investigations to assess their origin.  
  
The increased sensitivity of the O3 run allows us to set more stringent bounds on the binary merger rate density.  
The lack of a confirmed IMBH population as well as possible formation channels of IMBH distinct from those of
stellar-mass BHs preclude us from using an overall mass model for the IMBH population. Thus, we confine all 
the upper limit studies to a suite of discrete points in the IMBH parameter space. 
We incorporate more detailed physics in selecting the suite of IMBH binary waveforms as compared to earlier 
upper limit studies. In~\citet{Abbott:2017iws} we simulated a limited set of discrete mass and aligned-spin 
binary waveforms in the first advanced detector observation data to obtain upper limits on merger rate. 
The study with the first two observation runs used the most realistic \ac{NR} simulation set
with aligned spins for the upper limit study \citep{Salemi:2019ovz}. The most recent stringent merger rate
upper limit is $0.2$ \si{Gpc^{-3}yr^{-1}}, for the equal mass binary system with a component mass of $100\,\Msun \,$and component spins of dimensionless magnitude $0.8$ aligned with the binary orbital angular momentum.  
Recently, \citet{Chandra:2020ccy} used IMBH binary systems with generically spinning BHs with total mass 
between $210-500\,\Msun$ and obtained a most stringent upper limit of $0.28$ \si{Gpc^{-3}yr^{-1}} for equal-mass
binaries with total mass of 210 $\Msun$.
% and $\chi_\text{p} \sim 0.4243$ and $\chi_\text{eff} \sim 0.4243$. 

Here, we use a suite of NR simulations of GW emission from IMBH binary system with generically spinning
BHs in order to estimate our search sensitivity over the O3 data.  
We place the most stringent 90\% merger rate upper limit on equal mass and aligned spin BH binary of total 
mass $200\,\Msun$ and with individual BH spins of $0.8$ as {$0.056$ \si{Gpc^{-3}yr^{-1}}}. The revised limit 
is a factor $\sim 3.5$  more stringent than that obtained with the first two observing runs.  
We also update the merger rate for systems compatible with the source parameters of GW190521, first estimated
in \citet{GW190521Aastro}, to {$0.08^{+0.19}_{-0.07}$ \si{Gpc^{-3}yr^{-1}}}, using the combined search 
method applied to simulated signals injected over the entire O3 data.

The paper organization is as follows: Sect.~\ref{sec:data} summarizes the data being used for the search. 
Sect.~\ref{sec:searches} summarizes the combined search approach from the results from three distinct 
IMBH binary search algorithms. Sect.~\ref{sec:SearchResults} discusses the search results and followup of
the most significant candidate events. Sect.~\ref{sec:rates} provides a detailed discussion about the 
\ac{NR} GW injection set used and the rate upper limits study including the updated rate on 
the most significant GW190521-like systems.

%% file: data.tex
\section{Data Summary}
\label{sec:data}

We carry out the analysis using O3 data from both LIGO detectors 
(LHO-LIGO Hanford Observatory and LLO-LIGO Livingston Observatory) and the Virgo detector. We condition the data in multiple steps before performing our search ~\citep{LIGOScientific:2019hgc}. The strain data,
recorded from each detector, are calibrated in near real-time to produce an online 
data set~\citep{Viets2018, Acernese:2018bfl}. A higher-latency offline calibration stage 
provides identification of systematic errors and calibration configuration 
changes~\citep{Sun:2020wke, Estevez:2020pvj}.
The analyses presented here use the offline recalibrated data 
from the LIGO detectors, and the Virgo detector's online data.
For this search, we consider 246.2 days, 254.1 days, and 250.8 days of observing-mode data 
from the Hanford, Livingston, and Virgo detector respectively. The joint observation time
for the full network of three detectors is 156.4 days.

We then linearly subtract spectral features of known instrumental origin using auxiliary 
witness sensors, i.e.,  sensors that indicate the presence of noise causing these features. 
The subtraction removes calibration lines in all detectors, as well as 
60\,Hz harmonics produced by power mains coupling in the LIGO detectors~\citep{Driggers:2018gii,Davis:2018yrz}. 
Low-frequency modulation of the power mains coupling also results in sidebands 
around the 60\,Hz line; we apply an additional non-linear noise subtraction to remove  
these sidebands~\citep{Vajente:2019ycy}.

Periods of poor data quality are marked using data quality flags separated into three 
categories~\citep{LIGOScientific:2019hgc,Fisher:2020,Davis:2021ecd}, which are used to exclude
time segments from different searches, as described below. Category 1 flags indicate times
when a detector is not operating or recording data in its nominal state; these periods are not 
analyzed by any search. Category 2 flags indicate periods of excess noise that 
are highly likely to be caused by known instrumental effects. The cWB and PyCBC searches use 
different sets of category 2 flags. The GstLAL search does not use category 2 flags, as 
discussed in Sect. \ref{sec:searches}. Category 3 flags are based on statistical correlations 
with auxiliary sensors. Of the analyses presented here, only the cWB search uses category 3 flags.

The candidate events in this paper are vetted in the same way as past 
\ac{GW} events~\citep{TheLIGOScientific:2016zmo,GWTC2}. This 
validation procedure identifies data quality issues such as 
non-stationary noise or glitches of instrumental origin appearing in the strain 
data. Auxiliary sensors that monitor the detectors and environmental noise are 
used to check for artifacts that may either have accounted for, or contaminated the candidate signal~\citep{Nguyen:2021ybi}. 
For candidate events that coincide with glitches, subtraction of the glitches from the strain data is 
performed if possible~\citep{Cornish:2014kda,Littenberg:2015kpb,Pankow:2018qpo}; otherwise recommendations 
are made to exclude the relevant time or frequency ranges from parameter estimation analyses. 
Validation assessments for individual candidate events are provided in Sect.\ref{sec:SearchResults} 
and Appendix~\ref{sec:otherEvents}. 

%% file: searches.tex
\section{Search methods}
\label{sec:searches}

In this section, we describe the analysis methods algorithms (pipelines) used to search
the \acs{LVC} data from O3 for IMBH binary merger signals.  Such signals have short
durations in the detectors' sensitive frequency band, typically less than 1\,s.  
Thus, methods for detection of generic short transient \ac{GW} events (bursts) may 
be competitive compared to search methods which use parameterized models of
the expected signals (templates) from binary coalescences \citep[e.g. ][]{Chandra:2020ccy}.  
As in the IMBH binary search of O1 and O2~\citep{Abbott:2017iws,Salemi:2019ovz}, we employ both 
generic transient search methods and modelled template searches.  We first describe the 
generic transient pipeline, cWB, in the configuration used here, and then the two templated 
pipelines, GstLAL and PyCBC, which have been adapted to maximize sensitivity to IMBH
binary mergers. 
We then summarize the method used to combine the search outputs into a single candidate list, 
and finally discuss selection criteria to distinguish IMBH binary candidates from the 
known heavy stellar-mass BBH population \citep{GWTC2,o3apop}. 

The output of a transient search algorithm or pipeline is a set of candidate events, each
with an estimated time of peak strain at the participating detector(s).\footnote{For black hole
binary mergers, this peak strain time is close to the formation of a common horizon.} 
Each event is also assigned a ranking statistic value, and its significance is
quantified by estimating the corresponding \ac{FAR}, which is the expected number per
time of events caused by detector noise that have an equal or higher ranking statistic value. 

The sensitivity of a search to a population of IMBH mergers can be evaluated by
adding simulated signals (injections) to real GW detector strain data and analyzing
the resulting data streams, to output the ranking statistic and  
% with the standard search pipeline, including the 
estimated \ac{FAR} that each simulated signal would be assigned if present in an actual search.
Specific simulation campaigns will be described in detail in Sect.~\ref{sec:rates} and 
sensitivity estimates from individual search pipelines are included in a public data release. 

\subsection{cWB model waveform independent search for IMBH binaries}
\label{sec:cwb}
cWB~\citep{Klimenko:2004qh, Klimenko:2005xv, Klimenko:2006rh, Klimenko:2011hz, Klimenko:2015ypf,Drago:2020kic} is a GW search that uses minimal assumptions on signal morphology to detect and reconstruct GW transients. The search identifies
coincident energy across the network of detectors to classify GW signals. The cWB search has been 
participating in the search for IMBH signals since Initial LIGO's fifth science run~\citep{Abadie_2012}. The algorithm uses a multi-resolution wavelet transform, known as the Wilson
Daubechies Meyer wavelet transform~\citep{Necula:2012zz}, to map the multi-detector data into the
time-frequency domain, as blocks of a fixed time-frequency area known as pixels. 
The algorithm selects pixels with excess energy above the expected noise fluctuation and groups 
them into clusters, referred to as candidate events.
The collection and clustering of pixels differ based on the target source \citep{Klimenko:2015ypf}. 
Each candidate event is
ranked according to its coherent \ac{SNR} statistic \citep{Klimenko:2015ypf}, which incorporates the estimated coherent energy and residual noise energy.
An additional threshold is applied to the network correlation which provides the 
measure on the event correlation across multiple detectors in the network. 
The cWB algorithm reconstructs the source
sky location and whitened signal waveforms using the constrained maximum likelihood method~\citep{Klimenko:2015ypf}.

We estimate the \ac{FAR} of a search event with time lag analysis: data from one or more detectors are time-shifted by more than 1\,s with respect to other detectors in the network, then cWB identifies events in this time-shifted data.  Since the time-shift is greater than the \ac{GW} time of flight between detectors, this analysis estimates the rate and distribution of false alarms. 
The analysis is repeated many times with different time-shifts, yielding a total analyzed background time $T_{\mathrm{bkg}}$.
For a given search event, the \ac{FAR} value is estimated as the number of background events with coherent SNR 
greater than the value assigned to the event, divided by $T_{\mathrm{bkg}}$. 

The model independent nature of cWB search makes it susceptible to incorrectly classifying
noise artifacts. We apply a series of signal-dependent vetoes based on the time-frequency morphology and energy distribution properties to remove spurious noisy transients. We tune the veto 
values based on the extensive simulation of IMBH binary signals \citep[see Appendix A of ][]{PhysRevD.100.124022}. We divide the cWB search for quasi-circular BBH signals into two separate configurations: high-mass search and low-mass search, depending on the central frequency $f_\mathrm{c}$ of the GW signal. 
For a compact binary merger signal, $f_\mathrm{c}$ is inversely proportional to the redshifted total mass $M_z = (1+z)M$, where $M$ is the source frame total mass and $z$ is the source redshift. We then optimize the low-mass search sensitivity for signals with $f_\mathrm{c} > 
80\,$Hz (the BBH regime), and the high-mass search sensitivity for signals with $f_\mathrm{c} < 80\,$Hz (the IMBH regime).
In practice, a cut $f_\mathrm{c} > 60\,$Hz is imposed in the low-mass search and $f_\mathrm{c}< 100\,$Hz in the high-mass search, resulting in an overlap region covering $60 - 100 \,$Hz. In the O3 search, we combine the two searches by applying a trials factor of $2$ to the estimated \ac{FAR} 
for events in the overlap region. This improves the overall search sensitivity to borderline IMBH events~\citep{cwb_gw190521}.

The cWB search analyzes data from all three detectors in low latency. However, the follow up offline cWB 
analysis does not improve detection efficiency with the inclusion of Virgo. This is primarily due to 
the additional noise in the Virgo detector. Thus, at a given time, the cWB search uses the best available (most sensitive) two detector network configuration.
This ensures that the cWB search does not analyze the same data with multiple detector configurations.
In case, if any event shows high significance in 
low latency cWB analysis with the three-detector network and low significance in offline cWB analysis with 
the best two-detector configuration, we re-analyze that observing time with both the LLO-LHO-Virgo 
and LLO-LHO networks and apply a trials factor of $2$ to the minimum FAR over the two networks for the 
final significance.% \ac{FAR} estimate.

In the special case where an event shows high significance in low latency cWB analysis with the three-detector network and low significance in offline cWB analysis with the best two-detector configuration, we re-analyze the event with both the LLO-LHO-Virgo and LLO-LHO networks and apply a trials factor of $2$ to the minimum FAR over the two networks to establish its final significance.

\subsection{Templated searches for IMBH mergers}

For \ac{GW} signals whose forms are known or can be theoretically predicted, search
sensitivity is optimized by the use of matched filter templates that suppress 
noise realizations inconsistent with the predicted signals.  Since the binary 
parameters are \textit{a priori} unknown, a discrete set (bank) of templates is used
in order to cover signal parameter values within a predetermined range with a specified
minimum waveform accuracy 
\citep{Sathyaprakash:1991mt,Owen:1995tm}.  General binary black hole coalescence 
signals bear the imprint of component spins misaligned with the orbital axis, causing 
orbital precession, and potentially also of orbital eccentricity.  It is a so far 
unsolved problem to implement an optimal search over such a complex space of signals.

Instead, the searches presented here restrict the signal model to the dominant mode
of GW emission from quasi-circular, non-precessing binaries~\citep{Ajith:2009bn}, 
i.e.\ with component spins perpendicular to the orbital plane.  Both the GstLAL and PyCBC
searches use the SEOBNRv4 waveform approximant \citep{Bohe:2016gbl} as template waveforms, 
implemented as a reduced-order model \citep{Purrer:2015tud} for computational speed. 
These templates may still have high matches to signals from precessing or eccentric 
binaries, however in general, sensitivity to such signals will be reduced due to 
lower matches with template waveforms.

Each detector's strain time series is then correlated with each template to produce 
a matched filter time series.  Single-detector candidates are generated by identifying
maxima of the matched filter SNR above a predetermined threshold value.
However, during times of known disturbances in detector operation, or during very
high amplitude non-Gaussian excursions in the strain data, candidates are either not
produced or are discarded, since such high-\ac{SNR} maxima are very likely to be artifacts.
Signal consistency checks such as chi-squared \citep{Allen:2004gu} are also calculated 
and single-detector candidates may also be discarded for excessive deviation from 
the expected range of values.

If two or more
detectors are operating, their single-detector candidates are compared in order
to identify multi-detector candidate events which are consistent in the template
parameters, time of arrival, amplitude, and waveform phase over the detector network.  
The resulting multi-detector events are then ranked via a statistic which depends 
on the properties of single-detector candidates and their consistency %of those properties 
over the network.  Finally, the statistical
significance of each multi-detector event is obtained by comparing its statistic
value to the distribution expected for noise events, resulting in an estimate of
its \ac{FAR}.

In what follows we briefly summarize the methods specific to each of the matched filter pipelines.

\subsubsection{GstLAL search}
\label{GstLAL Search}
The search for IMBH mergers executed by the matched filter based GstLAL 
pipeline~\citep{gstlalmethods1, Chaddtdphi, sachdev2019gstlal} uses a template bank 
covering a parameter space of binaries with (redshifted) total masses in the range [50, 600]\,\Msun. The
mass ratios, $q=m_2/m_1$, of the binary systems covered lie between 1 and $1/10$, while their spins are
either aligned or anti-aligned with the total angular momentum of the system, with
the dimensionless spin magnitude less than 0.98. The analysis starts at a frequency
of 10\,Hz.

The SNR threshold applied for single-detector triggers is 4 for the Hanford and Livingston
detectors and 3.5 for the Virgo detector. The GstLAL search pipeline applies a signal-consistency test
based on the template's autocorrelation over time.  
%to assess the difference between the measured SNR time series and that predicted using the real signal. 
The search also uses a signal model to describe the prior probability of a binary from
a given source population being detected by each template: the signal model used for 
this search is uniform in the log of the reduced mass of the binary.

The ranking statistic applied to candidate events is an estimate of the relative
probability of the event's parameters being caused by a \ac{GW} signal as compared
to noise, i.e.\ the likelihood ratio. 
In addition to events formed from triggers from multiple detectors, triggers
found in a single detector are also included in the search, albeit with a penalty applied
to their ranking to account for the higher probability of noise origin.

The GstLAL search does not use data quality based vetoes of category 2 and above. 
Instead, the search uses data quality information known as iDQ~\citep{essick2020idq, godwin2020incorporation}, from
auxiliary channels monitoring the detector to compute a penalty term in the denominator of the 
ranking statistic. This has been computed for both single-detector and multi-detector 
triggers found by the search. The non-coincident and noise-like triggers are then used to estimate the
background noise probability density, which is sampled to find the estimated \ac{FAR}, corresponding to the likelihood ratio for a candidate~\citep{gstlalmethods1, sachdev2019gstlal}.

\subsubsection{PyCBC search}
\label{PyCBC Search}
The PyCBC-IMBH search used here \citep{PyCBC_Heavy} covers 
a target space of redshifted total masses between $100$ and $600\,M_\odot$,
with component masses greater than $40\, M_\odot$ and mass ratio between $1/1$ and $1/10$. 
The components have dimensionless spins projected onto the orbital axis between $-0.998$ and 
$0.998$. 
To reduce false alarms
arising from short-duration noise transients~\citep{Cabero:2019orq}, we discard any templates
with a duration less than $0.07$\,s, measured from the fixed starting frequency of $15$\,Hz.

The analysis 
pre-processes the data from each detector by windowing out very high amplitude excursions 
%(SNR $\gtrsim$ 100 or corresponding to 
($> 50 \sigma$ deviation from Gaussian noise) in the whitened strain 
time-series~\citep{Usman:2015kfa}. This gating step significantly 
suppresses the noise background. 
The \ac{SNR} threshold for trigger generation is chosen as $4$; any triggers in time marked 
by category-2 data quality veto are discarded.  We also remove LIGO triggers within 
$(-1, +2.5)$\,s of the centre of a gating window, since empirically such times contain many
lower-amplitude noise transients correlated with the central high amplitude glitch 
\citep{PyCBC_Heavy}. 

Signal-consistency $\chi_r^2$ and sine-Gaussian discriminant tests are applied to the remaining
triggers \citep{Allen:2004gu,Nitz_2018}. 
The single detector \acp{SNR} are corrected for short-term variation in the detector 
\ac{PSD} \citep{Nitz:2019hdf,Mozzon:2020gwa}, and a penalty is applied to 
triggers with a short-term PSD measure over $10$ times the expectation from stationary noise~\citep{PyCBC_Heavy}. 
The analysis also penalizes triggers with $\chi_r^2$ values above $10$, where the expectation for 
a well-matched signal is unity. These vetoes significantly reduce the background.

The search identifies candidates by checking the consistency between triggers in 2 or 3 detectors.
The resulting candidates are
ranked by using the expected distribution of astrophysical signal \acp{SNR}, phases and 
times over multiple detectors, as well as models of the non-Gaussian noise 
distribution in each template and detector \citep{Nitz:2017svb,Davies:2020tsx}. 
A \ac{FAR} is assigned to each candidate event by simulating the background noise distribution 
using time-shifted analyses~\citep{Usman:2015kfa}, similar to cWB.  The \acp{FAR} for events 
involving different detector combinations are finally combined as in \citep{Davies:2020tsx}. 

Other PyCBC-based searches overlapping the \ac{IMBH} parameter region were recently presented in 
\citet{Nitz:2019hdf, GWTC2} using two strategies: a broad parameter space search covering 
compact binaries from \ac{BNS} up to \ac{IMBH}, and a ``focused'' search for \ac{BBH} covering 
a restricted range of masses and with strict cuts to suppress noise artifacts.  The sensitivity
of the present IMBH search, at a \ac{FAR} threshold of $0.01\,\mathrm{yr}^{-1}$, 
to a set of simulated generically spinning 
 binary merger signals 
is increased relative to the broad (BBH) PyCBC searches of \citet{GWTC2} by a factor of $\sim 1.5$ 
($\sim 1.1$) in volume time (VT) for redshifted total mass $M_z \in [100-200]\,\Msun$, up
to a factor $\sim 2.8$ ($\sim 12.6$) for $M_z \in [450,600]\,\Msun$~\citep{PyCBC_Heavy}. 

\subsection{Combined search}

Each of our three targeted searches produces its list of candidates characterized by
GPS times and \ac{FAR} values.  The p-value for each candidate in a given search, defined
as the probability of observing one or more events from the noise alone with a detection
statistic as high as that of the candidate is then
\begin{equation}
    p = 1 - e^{-T\cdot \mathrm{FAR}},
\end{equation}
where $T$ is the total duration of data analyzed by the search.
We combine these lists to form a single list of candidates by first checking whether any 
events from different searches fall within a 0.1s time window of each other, 
and if so, selecting only the event with the lowest p-value $p_{\rm min}$.  The resulting
clustered events are ranked by a combined p-value,
\begin{equation}
    \Bar{p} \equiv 1 - (1 - p_{\rm min})^m,
\end{equation}
where $m$ denotes the trials factor (look-elsewhere) factor \citep{Abbott:2017iws,Salemi:2019ovz}.  
We take $m=3$ under the assumption that the noise backgrounds of our searches are
independent of one another.
If there is any correlation between these backgrounds, the effective trials
factor will be lower, which makes $m=3$ a conservative choice.

\subsection{Selection of Intermediate Mass Black hole Binaries }
\label{sec:imbh_definition}

As noted in~\citet{GWTC2, o3apop}, LIGO-Virgo observations include a population
of black hole binaries with component masses extending up to 60\,\Msun\ or above, and
remnant masses extending up to $\sim 100\,\Msun$; 
thus, there is \textit{a priori} no clear separation between such heavy \ac{BBH} systems
and the lightest \ac{IMBH} binaries.  We also expect search pipelines tuned for sensitivity
to \ac{IMBH} mergers to be capable of detecting such heavy stellar-mass \ac{BBH}, since the
overlap of their GW signals with those of \ac{IMBH} binaries may be large.  We find indeed that
many such BBH systems occurring within O3a are recovered with high significance by our
search. 

The complete catalog of such heavy BBH systems over the O3 run will be provided in a
subsequent publication, as an update to GWTC-2. 
Here, we select only those events for which, under the assumption that the signals were
produced by a quasi-circular binary black hole merger, we have clear evidence that
the remnant is an \ac{IMBH} of mass above 100\,\Msun, and at least
the primary black hole has a mass greater than the lower bound of the pair-instability mass gap. 
Strong evidence that this is the case for the primary component of GW190521 was presented
in~\citet{GW190521Adiscovery,GW190521Aastro}.

The selection criteria are evaluated as follows. We begin by defining the hypothesis $H$ according to which the detector 
output time series $d(t)$ is given by
\begin{equation}
d(t) = n(t) + h(t;\theta)
\end{equation}
where $n(t)$ is the noise time series, taken as a realisation of a zero-mean wide-sense stationary stochastic process, 
and $h(t;\theta)$ is the gravitational wave signal model dependent on a set of parameters $\theta$. 
We estimate the parameters $\theta$ by computing their posterior probability distribution
$p(\theta|d,H)$ using Bayes theorem:
\begin{equation}
p(\theta|d,H) = p(\theta|H)\frac{p(d|\theta,H)}{p(d|H)}
\end{equation}
where $p(\theta|H)$ is the prior probability distribution, $p(d|\theta,H)$ is the likelihood function -- taken as Normal distribution in the frequency domain with variance given by the power spectral density of the data $d(t)$ thanks to the wide-sense stationarity assumption -- and 
\begin{equation}
p(d|H) = \int d\theta\, p(\theta|H)p(d|\theta,H)
\end{equation}
is the \emph{evidence} for the hypothesis $H$. The latter quantity is particularly useful in the context of model selection. We can, in fact, compare the evidence for the signal hypothesis $H$ with the evidence for the hypothesis $N$ according to which no signal is present to assess their relative likelihoods by computing the ($\log_{10}$) Bayes factor
\begin{equation}\label{eq:bayes_factor}
\log_{10} B_{\rm SN} = \log_{10}\frac{p(d|H)}{p(d|N)}\,.
\end{equation}

Each potential candidate is followed up 
with a coherent Bayesian parameter estimation analysis~\citep{Veitch:2014wba,Lange:2017wki,Wysocki:2019grj}.
In these analyses, we model the GW signal as represented by precessing quasi-circular
waveforms from three different families: $\texttt{NRSur7dq4}$~\citep{Varma:2019csw}, $\texttt{SEOBNRv4PHM}$~\citep{Ossokine:2020kjp}
and $\texttt{IMRPhenomXPHM}$~\citep{Pratten:2020ceb}. All considered models include the effects of 
higher-order multipole moments as well as orbital precession due to misaligned \ac{BH} spins.  
Details of the analysis configuration follow previously published ones~\citep{GWTC2} and are 
documented in a separate paper~\citep{GWTC3}.  In particular, we consider uniform priors on the redshifted component masses, 
the individual spin magnitudes, and the luminosity distance proportional to its square modulus. For the source 
orientation and spin vectors, we employ isotropic priors.

As a quantitative criterion to select a GW event as an \ac{IMBH} binary,
we consider the support of the joint posterior distributions for the primary mass $m_1$ and of the remnant
mass $M_\textrm{f}$.  For reference values 
$M^\ast = 100 \Msun$ and $m_1^\ast = 65 \Msun$, we label
a candidate an intermediate-mass black hole binary if
\begin{equation}\label{eq:p-imbh}
    \int^{+\infty}_{M^\ast} \int^{+\infty}_{m_1^\ast} dm_1\,dM_\textrm{f}~p(M_\textrm{f},m_1|D,H) \geq p^\ast,
\end{equation}
where $p^\ast$ is a reference probability threshold, chosen to be $p^\ast = 0.9$. 
To perform the integral in Eq.~(\ref{eq:p-imbh}), we 
construct a Gaussian kernel density estimate to interpolate the posterior $p(M_\textrm{f},m_1|D,H)$ 
which we use to perform the integral on a grid.

Thus, the main list of candidates presented in the following section does not correspond 
to the complete set of events recovered by the searches, but only to those relevant 
to a potential astrophysical IMBH population.  However, for comparison with earlier 
results \citep{GWTC2}, we also report a full list of events detected by the combined IMBH search 
in O3a data, including BBH events that do not fall into the IMBH region: see Sect.~\ref{sec:o3a_bbh}.

The data for some events may not be consistent with the quasi-circular BBH
signal plus Gaussian noise model, either because they contain a signal which deviates
significantly from this standard BBH model, or are affected by detector noise artefacts
that cannot be removed or mitigated.  In such cases the values of $p(M_\textrm{f}|D,H)$ and 
$p(m_1|D,H)$ extracted from the Bayesian analysis may either be inaccurate or indeed 
meaningless, for events arising from instrumental noise or even from a putative
astrophysical source that is not a compact binary merger.  Such events
will \emph{not} be excluded from results presented here: they
will be individually discussed in the following sections. 

%% file: results.tex
\section{Search results}
\label{sec:SearchResults}

\begin{table*}
\centering
% scalebox not necessary?
{
\begin{tabular}{llcccc}
\toprule
    Events &      GPS Time &             cWB FAR (yr$^{-1}$) &           PyCBC FAR (yr$^{-1}$) &          GstLAL FAR (yr$^{-1}$) &        $\Bar{p}$ \\
\hline
  GW190521 &  1242442967.5 & $2.0\times 10^{-4}$ & $1.4\times 10^{-3}$ & $1.9\times 10^{-3}$ & $4.5\times 10^{-4}$ \\
  \januaryevent\,$^\dag$ &  1263002916.2 & $5.8\times 10^{-2}$ & $8.6\times 10^{+2}$ & $3.6\times 10^{+4}$ & $1.2\times 10^{-1}$ \\
 \valentinesdaynoise &  1265755544.5 & $1.3\times 10^{-1}$ &                   - &                   - & $2.5\times 10^{-1}$ \\
\hline
\vspace*{1mm}
\end{tabular}
}
\caption{\label{Table:candidates} Events from the combined search for intermediate mass black hole
binary mergers in O3 data, sorted by their combined p-value $\Bar{p}$. 
$^\dag$\,\januaryevent~was recovered by the cWB search using LHO-LLO data with 
a \ac{FAR} of 15.87\,yr$^{-1}$ and by a followup search using LHO-LLO-Virgo data with a \ac{FAR} of 
0.029\,yr$^{-1}$; the FAR quoted in the table for cWB is derived from the LHO-LLO-Virgo search with a trials
factor of 2.}
\end{table*}

\subsection{Candidate IMBH events}
\label{sec:mainlist}

The individual searches are applied on the full O3 data with the analysis time of 0.734\,yr, 0.747\,yr and 0.874\,yr for cWB, PyCBC-IMBH and GstLAL-IMBH search respectively. 
Table~\ref{Table:candidates} summarises the results from the combined cWB-GstLAL-PyCBC IMBH
search on full O3 data detailed in Sect.\ref{sec:searches}. These events have a combined p-value less than 0.26 (a threshold determined by the 
loudest noise event in the combined search, \valentinesdaynoise~and satisfy the criteria
for potential IMBH binary sources of Sect.~\ref{sec:imbh_definition}. 
For completeness, we have also listed marginal triggers found by our combined search in Appendix~\ref{sec:otherEvents}. 

The top-ranked event 
is GW190521 and it has a highly significant combined p-value of $4.5 \times 10^{-4}$. 
If this signal is from a quasi-circular merger, then the signal is found to be consistent with the merger of two black holes 
in a mildly precessing orbit, with component masses of $85^{+21}_{-14} \Msun$ and $66^{+17}_{-18} \Msun$ and a remnant black hole of $142^{+28}_{-16} \Msun$ falling in the mass range of 
intermediate-mass black holes. 
A full description of GW190521 and its implications can be found in \citet{GW190521Adiscovery,GW190521Aastro}.

The second-ranked candidate, \januaryevent, was observed on 14th January 2020 at 02:08:18 UTC and identified by the low 
latency cWB search in the LHO-LLO-Virgo detector network configuration, with a \ac{FAR} of $< 0.04\,\mathrm{yr}^{-1}$. The event was publicly reported via GCN minutes after the event was observed~\citep{LIGOPublicGCNs}. 
Given the significance of the low-latency alert with the 3-detector configuration, 
we employ both LHO-LLO and LHO-LLO-Virgo networks in cWB to estimate the significance for this event: we find 
\acp{FAR} of $15.87\,\mathrm{yr}^{-1}$ and $0.029\,\mathrm{yr}^{-1}$ for these configurations, respectively. The SNR reconstructed by cWB 
for each network configuration is 12.3 and 14.5, respectively.
As mentioned above in \ref{sec:cwb}, we apply a
trials factor of 2 to the most significant result, 
obtaining a \ac{FAR} of $0.058\,\mathrm{yr}^{-1}$ for the cWB search. 
The combined p-value of this event, $0.12$, is marginally significant. 

We then examined possible environmental or instrumental causes for the candidate signal.  
Excess vibrational noise could have contributed to the signal in the LIGO Hanford detector, as 
discussed in Appendix~\ref{sec:S200114fnoise}. 
Furthermore, the morphology of \januaryevent~is consistent with a well-studied class of glitches
known as Tomtes~\citep{Buikema:2020dlj,Davis:2021ecd}, which occur multiple times per hour
in LIGO Livingston. However, we are not currently able to exclude 
a putative morphologically similar astrophysical signal, as there are no known 
instrumental auxiliary channels that couple to this glitch type. 
We undertake detailed model-independent event reconstruction and parameter estimation (PE) studies, 
summarized in Appendix~\ref{sec:200114}. 
Although model independent methods/algorithms produce mutually consistent reconstructions of the event, 
our analysis using the available quasi-circular BBH merger waveforms does not support a 
consistent interpretation of the event as a binary merger signal present across the detector 
network.
We cannot conclusively rule out an astrophysical origin for the event, however it also appears 
consistent with an instrumental artefact in LLO in coincidence with noise fluctuations in LHO
and Virgo.

The third-ranked event was observed by the cWB pipeline on 14th February, 2020 
at 22:45:26 UTC with a combined SNR of 13.1 in the two Advanced LIGO Detectors. The event has a 
$p_{\rm{cWB}}=0.092$ and thus a $\Bar{p}=0.251$.  In addition to its marginal significance, 
the event has characteristics consistent with an instrumental noise transient. 
Excess noise due to fast scattered light~\citep{Soni:2021cjy} is present in both LLO and LHO data. 
At Livingston, the excess noise extends up to 70\,Hz and lasts many seconds before and after the event. 
The Hanford scattering noise is weaker in amplitude but still overlaps completely with the duration of the event.
Since it is the most significant noise event obtained in the combined search with the 
cWB pipeline in its production configuration considering only 2-detector events, we use \valentinesdaynoise~to
establish a threshold of significance for inference of \ac{IMBH} merger rates (for which see Sect.~\ref{sec:rates}).
As \valentinesdaynoise~is likely caused by detector noise, any events with lower significance may be assumed to have 
a high probability of noise origin.  For completeness, we discuss some marginal events from the combined search
in Appendix \ref{sec:otherEvents}.

\begin{table*}[tbh]
  \begin{center}
  \scalebox{0.9}{
\begin{tabular}{c|c|ccc|cc|c}
\toprule
      & {cWB} & \multicolumn{3}{c|}{PyCBC} & \multicolumn{2}{c|}{GstLAL} & Combined  \\ \hline
      &       & GWTC-2 Broad & GWTC-2 BBH  & IMBH & GWTC-2 Broad & IMBH       & IMBH \\
 Event & {FAR (yr$^{-1}$)} & \multicolumn{3}{c|}{FAR (yr$^{-1}$)} & \multicolumn{2}{c|}{FAR (yr$^{-1}$)} & $\Bar{p}$  \\ \hline
GW190408\_181802 &  $9.5\times 10^{-4}$ & $<2.5\times 10^{-5}$ & $<7.9\times 10^{-5}$ & $1.6\times 10^{-2}$ & $<1.0\times 10^{-5}$ & $<1.0\times 10^{-5}$ & $ < 1.0 \times 10^{-4}$ \\
 GW190413\_052954 &                    - &                    - &  $7.2\times 10^{-2}$ & $5.6\times 10^{-1}$ &                    - &  $5.4\times 10^{+3}$ &     $7.1\times 10^{-1}$ \\
 GW190413\_134308 &                    - &                    - &  $4.4\times 10^{-2}$ & $1.4\times 10^{-1}$ &  $3.8\times 10^{-1}$ &  $1.2\times 10^{+3}$ &     $2.7\times 10^{-1}$ \\
 GW190421\_213856 &  $3.0\times 10^{-1}$ &  $1.9\times 10^{+0}$ &  $6.6\times 10^{-3}$ & $6.1\times 10^{-3}$ &  $7.7\times 10^{-4}$ &  $1.8\times 10^{+0}$ &     $1.4\times 10^{-2}$ \\
 GW190503\_185404 &  $1.8\times 10^{-3}$ &  $3.7\times 10^{-2}$ & $<7.9\times 10^{-5}$ & $2.5\times 10^{-3}$ & $<1.0\times 10^{-5}$ &  $1.7\times 10^{-1}$ &     $4.0\times 10^{-3}$ \\
 GW190512\_180714 &  $8.8\times 10^{-3}$ &  $3.8\times 10^{-5}$ & $<5.7\times 10^{-5}$ & $4.0\times 10^{+1}$ & $<1.0\times 10^{-5}$ & $<1.0\times 10^{-5}$ & $ < 1.0 \times 10^{-4}$ \\
 GW190513\_205428 &                    - &  $3.7\times 10^{-4}$ & $<5.7\times 10^{-5}$ & $5.0\times 10^{-2}$ & $<1.0\times 10^{-5}$ &  $2.1\times 10^{-1}$ &     $1.1\times 10^{-1}$ \\
 GW190514\_065416 &                    - &                    - &  $5.3\times 10^{-1}$ & $1.1\times 10^{+0}$ &                    - &  $7.6\times 10^{+2}$ &     $9.2\times 10^{-1}$ \\
 GW190517\_055101 &  $8.0\times 10^{-3}$ &  $1.8\times 10^{-2}$ & $<5.7\times 10^{-5}$ & $8.7\times 10^{-4}$ & $9.6\times 10^{-4}$ &  $2.7\times 10^{-2}$ &     $1.9\times 10^{-3}$ \\
 GW190519\_153544 &  $3.1\times 10^{-4}$ & $<1.8\times 10^{-5}$ & $<5.7\times 10^{-5}$ & $<1.1\times 10^{-4}$ & $<1.0\times 10^{-5}$ &  $3.9\times 10^{-3}$ &     $2.5\times 10^{-4}$ \\
         GW190521 &  $2.0\times 10^{-4}$ &  $1.1\times 10^{+0}$ &                    - & $1.4\times 10^{-3}$ &  $1.2\times 10^{-3}$ &  $1.9\times 10^{-3}$ &     $4.5\times 10^{-4}$ \\
 GW190521\_074359 & $<1.0\times 10^{-4}$ & $<1.8\times 10^{-5}$ & $<5.7\times 10^{-5}$ & $ < 2.3\times 10^{-5}$ & $<1.0\times 10^{-5}$ & $<1.0\times 10^{-5}$ & $ < 1.0 \times 10^{-4}$ \\
 GW190602\_175927 &  $1.5\times 10^{-2}$ &                    - &  $1.5\times 10^{-2}$ & $1.1\times 10^{-3}$ &  $1.1\times 10^{-5}$ & $<1.0\times 10^{-5}$ & $ < 1.0 \times 10^{-4}$ \\
 GW190701\_203306 &  $3.2\times 10^{-1}$ &                    - &                    - & $<1.9\times 10^{-4}$ &  $1.1\times 10^{-2}$ &  $3.8\times 10^{-2}$ &     $4.3\times 10^{-4}$ \\
 GW190706\_222641 & $<1.0\times 10^{-3}$ &  $6.7\times 10^{-5}$ &  $4.6\times 10^{-5}$ & $<1.1\times 10^{-4}$ & $<1.0\times 10^{-5}$ &  $2.4\times 10^{-3}$ &     $2.5\times 10^{-4}$ \\
 GW190727\_060333 &  $8.8\times 10^{-2}$ &  $3.5\times 10^{-5}$ &  $3.7\times 10^{-5}$ & $<1.2\times 10^{-4}$ & $<1.0\times 10^{-5}$ &  $4.5\times 10^{-4}$ &     $2.7\times 10^{-4}$ \\
 GW190731\_140936 &                    - &                    - &  $2.8\times 10^{-1}$ & $6.4\times 10^{-1}$ &  $2.1\times 10^{-1}$ &  $2.1\times 10^{+0}$ &     $7.6\times 10^{-1}$ \\
 GW190803\_022701 &                    - &                    - &  $2.7\times 10^{-2}$ & $1.7\times 10^{-1}$ &  $3.2\times 10^{-2}$ &  $3.0\times 10^{+0}$ &     $3.2\times 10^{-1}$ \\
 GW190828\_063405 & $<9.6\times 10^{-4}$ & $<1.0\times 10^{-5}$ & $<3.3\times 10^{-5}$ & $<7.0\times 10^{-5}$ & $<1.0\times 10^{-5}$ & $<1.0\times 10^{-5}$ & $ < 1.0 \times 10^{-4}$ \\
 GW190915\_235702 & $<1.0\times 10^{-4}$ &  $8.6\times 10^{-4}$ & $<3.3\times 10^{-5}$ & $3.8\times 10^{-4}$ & $<1.0\times 10^{-5}$ &  $4.7\times 10^{-1}$ &     $2.2\times 10^{-4}$ \\
 GW190929\_012149 &                    - &                    - &                    - & $3.1\times 10^{-1}$ &  $2.0\times 10^{-2}$ &  $2.9\times 10^{+1}$ &     $5.0\times 10^{-1}$ \\
 \hline
\end{tabular}}
\vspace*{1mm}
 \caption{Candidate events from this search for \ac{IMBH} mergers in O3a data, including 
 binary black hole mergers outside the \ac{IMBH} parameter space, and comparison with 
 previously obtained GWTC-2 results from the templated search algorithms~\citep{GWTC2}.  
 The cWB search algorithm used here is unchanged over GWTC-2.
 Candidates are sorted by GPS time and the \ac{FAR} is provided for each search algorithm.  
 Templated methods used in GWTC-2 comprise the PyCBC and GstLAL broad parameter 
 space pipelines and the PyCBC BBH-focused pipeline, while the optimized algorithms
 applied in this search are labelled ``IMBH''. 
 The event names encode the UTC date with the time of the event given after the underscore, 
 except for the individually published event GW190521. The GstLAL FAR values have been capped 
 at $1.0 \times 10^{-5}$ \si{yr^{-1}} and corresponding $\Bar{p}$ values have also been capped. 
 For PyCBC events with \ac{FAR} estimates limited by finite background statistics, an upper
 limit is stated.  The IMBH combined search p-values $\Bar{p}$ for each event are calculated from 
 Eq.~\eqref{eq:p-imbh} using p-values of the cWB, PyCBC-IMBH and GstLAL-IMBH searches.% for that event.
 For details of the search configurations and event parameters, refer to~\citet{GWTC2}.}
  \label{table:O3aCBC}
  \end{center}
\end{table*}

\subsection{Complete O3a search results including BBH}
\label{sec:o3a_bbh}

As noted earlier in ~\ref{sec:searches}, 
the template-based searches have high sensitivity to the known population of heavy 
stellar-mass BBH mergers, which may be compared to 
searches deployed in GWTC-2~\citep{GWTC2}.
Here 
we record the complete list of significant events recovered
by the combined IMBH search from O3a data in Table~\ref{table:O3aCBC}, and supply  
corresponding search results from GWTC-2 for comparison. 
Specifically, we show outputs from the PyCBC broad parameter space and focused BBH 
searches~\citep{Nitz:2019hdf} and the GstLAL broad parameter space search~\citep{sachdev2019gstlal}. 

For GW190521, the PyCBC IMBH search yields a \ac{FAR} of $1.4\times 10^{-3}\,\mathrm{yr}^{-1}$,
as compared to $1.1\,\mathrm{yr}^{-1}$ for the broad parameter space analysis of \citet{GW190521Adiscovery,GWTC2}. 
This significant change is in part because the PyCBC IMBH search is optimized 
for shorter duration signals, and does not consider potential signals of total mass 
significantly below 100\,\Msun; the mass and spin values of GW190521 are also 
likely not covered by the templates used in earlier PyCBC searches, which imposed 
a minimum duration of $0.15$\,s. 
A similar change in statistical significance is also
observed for GW190602$\_175927$ for the same reasons. However, the IMBH search 
results assign lower significance to GW190519$\_153544$ and GW190706$\_222641$ as 
compared to the GWTC-2 results. 

The GstLAL pipeline recovers the GW190521 event at a \ac{FAR} of $1.9\times 10^{-3}\,\mathrm{yr}^{-1}$ 
over all of O3 data. It was reported earlier~\citep{GW190521Adiscovery,GWTC2} 
at a \ac{FAR} of $1.2\times 10^{-3}\,\mathrm{yr}^{-1}$ over O3a. As described in~\ref{GstLAL Search},
the GstLAL pipeline has employed a dedicated search for IMBH binaries with better coverage for the heavier mass binaries than the catalog search. Also, the iDQ
based data quality information used to inform the calculation of the ranking statistics,
now incorporates multi-detector triggers, as against the only single detector triggers
that were used before.  Differences in the significance of the events found by the IMBH 
specific GstLAL search presented here, with what was reported for O3a in~\citet{GWTC2}
can be attributed to the differences in the search settings and the data spanning over all of O3. 

%% file: rate.tex
\section{Astrophysical Rates of IMBH Binary Coalescence}
\label{sec:rates}

Improved detector sensitivity, updated search methods, and the detection of GW190521 allow us to obtain revised 
bounds on the merger rate (strictly, rate density) of IMBH binaries.  Due to the lack of knowledge of specific formation channels for 
IMBH binaries, even more so than for stellar-mass BH binaries, and the sparse observational evidence of any IMBH population,
we do not consider any overall mass model for such a population. Instead, here we simulate a suite of IMBH binary waveforms
for discrete points in parameter space, including generically spinning component \acp{BH}, derived from \ac{NR} simulations. 
A similar campaign was carried out in \citep{Salemi:2019ovz} using NR waveforms for IMBH binaries having component BH spins aligned with the binary orbital axis, injected into the O1 and O2 data.

\subsection{Injection Set}
Here, we report on the merger rate of IMBH binary sources based on NR simulations computed by the SXS~\citep{Mroue_2013},
RIT~\citep{Healy_2017}, and GeorgiaTech~\citep{Jani_2016} codes. These simulations include higher-order
multipoles, which may make important contributions to the detection of high-mass and low mass-ratio ($q \leq 1/4$) 
binaries~\citep{Bustillo:2015qty}. 
Based on previous studies which measured the agreement between different NR
codes~\citep{Salemi:2019ovz}, we include the following harmonic modes in our analysis:
$(\ell,m) = \{(2, \pm 1), (2, \pm 2), (3, \pm2), (3, \pm 3), (4, \pm 2), (4, \pm 3), (4, \pm 4)\}$.

We consider 43 IMBH binary sources with fixed source frame
masses and spins, shown in Table~\ref{Table:Injections}. These 43 sources include a subset of 16 sources 
investigated in the O1-O2 IMBH binary search.  This updated search includes sources with total mass up to  
$800\,\Msun$ and expands the range of targeted mass ratio $q$ to between $1/1 - 1/10$. 
We also further explore the effects of the component spins on detection efficiency.
Of the 43 targeted IMBH sources, 4 have spins aligned with the orbital axis, with effective total 
spin~\citep{Ajith:2009bn} $\chi_\mathrm{eff} \equiv (\chi_{1,\parallel} + q \chi_{2,\parallel})/(1+q) = 
0.8$, where $\chi_{\parallel}$ denotes the \ac{BH} spin
resolved along the orbital axis, and 4 have anti-aligned spins with $\chi_\mathrm{eff}=-0.8$.  A further 11 
have precessing spins: $\chi_\mathrm{p} \neq 0$, where $\chi_\mathrm{p}$ is the effective 
spin-precession parameter of \citep{Hannam:2013oca,Schmidt:2014iyl}.

The simulated signals for each targeted source point are uniformly distributed in sky location 
$(\theta, \phi)$ and inclination angle $\cos(\iota)$. 
The source redshift $z$ is uniformly distributed in comoving volume, according to the 
TT+lowP+lensing+ext cosmological parameters given in Table IV of Ref.~\citet{Ade:2015xua}, up to a maximum 
redshift $z_\mathrm{max}$.  The signals are added to the O3 strain data, i.e.\ injected, with a uniform 
spacing in time approximately every 100\,s over the full observing time, $T_0 = 363.38$ days.

To avoid generating injections that are well outside any possible detection range, $z_\mathrm{max}$ is 
calculated for each IMBH source point independently. 
We consider values of redshift $z$ in increments of 0.05 and calculate a conservative upper bound on the 
optimal three-detector network SNR, SNR$_\mathrm{net}$, for each $z$. To bound the optimal SNR in a single
detector, we assume the source is face-on $\cos(\iota) = 1$, located directly overhead the detector, and we estimate the 
detector's PSD using $\sim 8$ hours of typical O3 data.
For precessing waveforms, the $\iota$ is set at 10\,Hz.
We determine the maximum redshift by requiring that SNR$_\mathrm{net}(z_\mathrm{max}) \sim 5$. 
This results a range of $z_\mathrm{max}$ across all targeted sources from 0.05 for the $(400+400)\,\Msun$ 
anti-aligned spin source to 2.75 for the $(100+100)\,\Msun$ aligned-spin source, as in 
Table~\ref{Table:Injections}.

When generating the injection parameters, we impose an additional threshold SNR$_\mathrm{net} > 5$ to limit 
the number of simulations injected into detection pipelines that have a negligibly small probability of detection.
For this purpose, the SNR$_\mathrm{net}$ is re-estimated, taking into account the randomly selected source position
and orientation.  We thus assume simulated events with SNR$_\mathrm{net} < 5$ are missed by the search pipelines; 
these events are, though, accounted for in the calculation of sensitive volume and merger rates.  

As stated in Sect.~\ref{sec:searches}, the searches process the remaining injections with the same configuration
as used for results from O3 data.  This is necessary to obtain unbiased rate estimates.  In the case of cWB, 
injections were processed with the most sensitive two-detector configuration: thus, for consistency, we consider 
only events recovered in the corresponding offline two-detector search results. 

\begin{table*}[htb]
\centering
\scalebox{0.85}{
\begin{tabular}{c|c|c|c|c|c|c|c}
\hline
$M~(\Msun)$ & q & $\chi_{\rm{eff}}$ & $\chi_\mathrm{p}$ & SIM ID & $z_{\rm{max}}$ & $\langle VT \rangle_{\rm{sen}}$ [\si{Gpc^3yr}]
& $R_{90\%}$ [\si{Gpc^{-3}yr^{-1}}] 
\\
\hline
120 &  1/2 &    0.00 &  0.00 &  SXS:BBH:0169, RIT:BBH:0117:n140, GT:0446 &  2.00 &   12.42 &   0.19 \\
   120 &  1/4 &    0.00 &  0.00 &  SXS:BBH:0182, RIT:BBH:0119:n140, GT:0454 &  1.35 &    5.08 &   0.45 \\
   120 &  1/5 &    0.00 &  0.00 &  SXS:BBH:0056, RIT:BBH:0120:n140, GT:0906 &  1.15 &    3.45 &   0.67 \\
   120 &  1/7 &    0.00 &  0.00 &    SXS:BBH:0298 RIT:BBH:Q10:n173, GT:0568 &  0.90 &    1.85 &   1.24 \\
   120 & 1/10 &    0.00 &  0.00 &           SXS:BBH:0154, RIT:BBH:0068:n100 &  0.70 &    0.91 &   2.52 \\
   150 &  1/2 &    0.00 &  0.00 &  SXS:BBH:0169, RIT:BBH:0117:n140, GT:0446 &  1.85 &   12.84 &   0.30 \\
   200 &    1 &    0.00 &  0.00 &  SXS:BBH:0180, RIT:BBH:0198:n140, GT:0905 &  1.85 &   16.04 &   0.14 \\
   200 &  1/2 &    0.00 &  0.00 &  SXS:BBH:0169, RIT:BBH:0117:n140, GT:0446 &  1.60 &   11.67 &   0.20 \\
   200 &  1/4 &    0.00 &  0.00 &  SXS:BBH:0182, RIT:BBH:0119:n140, GT:0454 &  1.15 &    4.80 &   0.48 \\
   200 &  1/7 &    0.00 &  0.00 &    SXS:BBH:0298 RIT:BBH:Q10:n173, GT:0568 &  0.80 &    1.74 &   1.32 \\
   220 & 1/10 &    0.00 &  0.00 &           SXS:BBH:0154, RIT:BBH:0068:n100 &  0.60 &    0.81 &   2.86 \\
   250 &  1/4 &    0.00 &  0.00 &  SXS:BBH:0182, RIT:BBH:0119:n140, GT:0454 &  1.00 &    3.90 &   0.59 \\
   300 &  1/2 &    0.00 &  0.00 &  SXS:BBH:0169, RIT:BBH:0117:n140, GT:0446 &  1.15 &    7.55 &   0.31 \\
   350 &  1/6 &    0.00 &  0.00 &  SXS:BBH:0181, RIT:BBH:0121:n140, GT:0604 &  0.60 &    1.13 &   2.03 \\
   400 &    1 &    0.00 &  0.00 &  SXS:BBH:0180, RIT:BBH:0198:n140, GT:0905 &  1.00 &    5.65 &   0.41 \\
   400 &  1/2 &    0.00 &  0.00 &  SXS:BBH:0169, RIT:BBH:0117:n140, GT:0446 &  0.85 &    4.06 &   0.57 \\
   400 &  1/3 &    0.00 &  0.00 &  SXS:BBH:0030, RIT:BBH:0102:n140, GT:0453 &  0.70 &    2.55 &   0.90 \\
   400 &  1/4 &    0.00 &  0.00 &  SXS:BBH:0182, RIT:BBH:0119:n140, GT:0454 &  0.60 &    1.70 &   1.36 \\
   400 &  1/7 &    0.00 &  0.00 &    SXS:BBH:0298 RIT:BBH:Q10:n173, GT:0568 &  0.45 &    0.68 &   3.38 \\
   440 & 1/10 &    0.00 &  0.00 &                 RIT:BBH:Q10:n173, GT:0568 &  0.30 &    0.31 &   7.51 \\
   500 &  2/3 &    0.00 &  0.00 &                RIT:BBH:0115:n140, GT:0477 &  0.70 &    2.39 &   0.96 \\
   600 &    1 &    0.00 &  0.00 &  SXS:BBH:0180, RIT:BBH:0198:n140, GT:0905 &  0.55 &    1.09 &   2.12 \\
   600 &  1/2 &    0.00 &  0.00 &  SXS:BBH:0169, RIT:BBH:0117:n140, GT:0446 &  0.50 &    0.99 &   2.32 \\
   800 &    1 &    0.00 &  0.00 &  SXS:BBH:0180, RIT:BBH:0198:n140, GT:0905 &  0.35 &    0.20 &  11.76 \\
   200 &    1 &    0.80 &  0.00 &           SXS:BBH:0230, RIT:BBH:0063:n100 &  2.75 &   40.34 &   0.06 \\
   400 &    1 &    0.80 &  0.00 &           SXS:BBH:0230, RIT:BBH:0063:n100 &  1.55 &   20.07 &   0.11 \\
   600 &    1 &    0.80 &  0.00 &           SXS:BBH:0230, RIT:BBH:0063:n100 &  0.95 &    6.46 &   0.36 \\
   800 &    1 &    0.80 &  0.00 &           SXS:BBH:0230, RIT:BBH:0063:n100 &  0.65 &    1.36 &   1.70 \\
   200 &    1 &   -0.80 &  0.00 &            SXS:BBH:0154,RIT:BBH:0068:n100 &  1.45 &   11.40 &   0.20 \\
   400 &    1 &   -0.80 &  0.00 &            SXS:BBH:0154,RIT:BBH:0068:n100 &  0.75 &    2.33 &   0.99 \\
   600 &    1 &   -0.80 &  0.00 &            SXS:BBH:0154,RIT:BBH:0068:n100 &  0.40 &    0.29 &   7.88 \\
   800 &    1 &   -0.80 &  0.00 &            SXS:BBH:0154,RIT:BBH:0068:n100 &  0.25 &    0.06 &  38.27 \\
   200 &    1 &    0.51 &  0.42 &                                   GT:0803 &  2.15 &   27.72 &   0.08 \\
   200 &  1/2 &    0.14 &  0.42 &                                   GT:0872 &  1.90 &   15.45 &   0.15 \\
   200 &  1/4 &    0.26 &  0.42 &                                   GT:0875 &  1.55 &    9.20 &   0.25 \\
   200 &  1/7 &    0.32 &  0.42 &                                   GT:0888 &  1.15 &    4.30 &   0.54 \\
   400 &    1 &    0.51 &  0.42 &                                   GT:0803 &  1.20 &   11.79 &   0.20 \\
   400 &  1/2 &    0.14 &  0.42 &                                   GT:0872 &  1.05 &    6.45 &   0.36 \\
   400 &  1/4 &    0.26 &  0.42 &                                   GT:0875 &  0.90 &    4.28 &   0.54 \\
   400 &  1/7 &    0.32 &  0.42 &                                   GT:0888 &  0.70 &    2.12 &   1.08 \\
   600 &    1 &    0.51 &  0.42 &                                   GT:0803 &  0.70 &    3.02 &   0.76 \\
   600 &  1/2 &    0.14 &  0.42 &                                   GT:0872 &  0.60 &    1.73 &   1.33 \\
   800 &    1 &    0.51 &  0.42 &                                   GT:0803 &  0.45 &    0.22 &  10.28 \\
\hline
\end{tabular}}
\caption{\label{Table:Injections} Summary of the source frame parameters,
sensitive volume-time and merger rate density upper limit at 90\% confidence.  For the upper limit, we assume
no detection except for the non-spinning system with total mass $M_T=150\,\Msun$ and $q=1/2$ marked with $\dagger$, 
for which we have assumed one detection. The source spin parameters are defined at a starting frequency of 16\,Hz.
}
\end{table*}

\subsection{Sensitive Volume Time and Merger Rate}
% KC, TD
\begin{figure*}
    \hspace*{-1cm}
    \centering
    \includegraphics[scale=0.4]{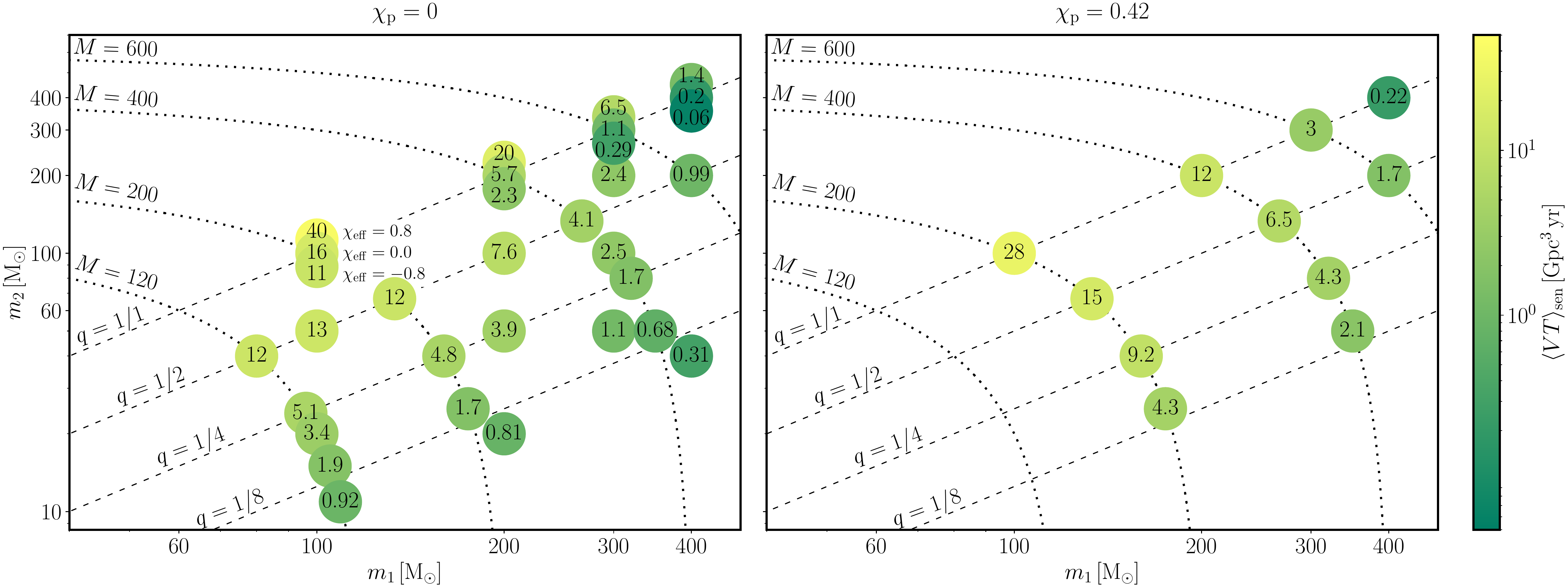}
    \caption{The averaged sensitive time-volume $\langle VT \rangle_{\textrm{sen}}$ in Gpc$^3\,$yr for the targeted \ac{IMBH} binary sources
    in the $m_1$-$m_2$ plane. The values are rounded where necessary for display. Left panel for $\chi_\mathrm{p}=0$ and right panel for $\chi_\mathrm{p}=0.42$. Each circle corresponds
    to one class of \ac{IMBH} binaries in the source frame. The $\chi_\textrm{eff}$ values of injection sets are labelled 
    and shown as displaced circles. }
    \label{Fig:VTsens}
\end{figure*}

\begin{figure*}
    \hspace*{-1cm}
    \centering
    \includegraphics[scale=0.4]{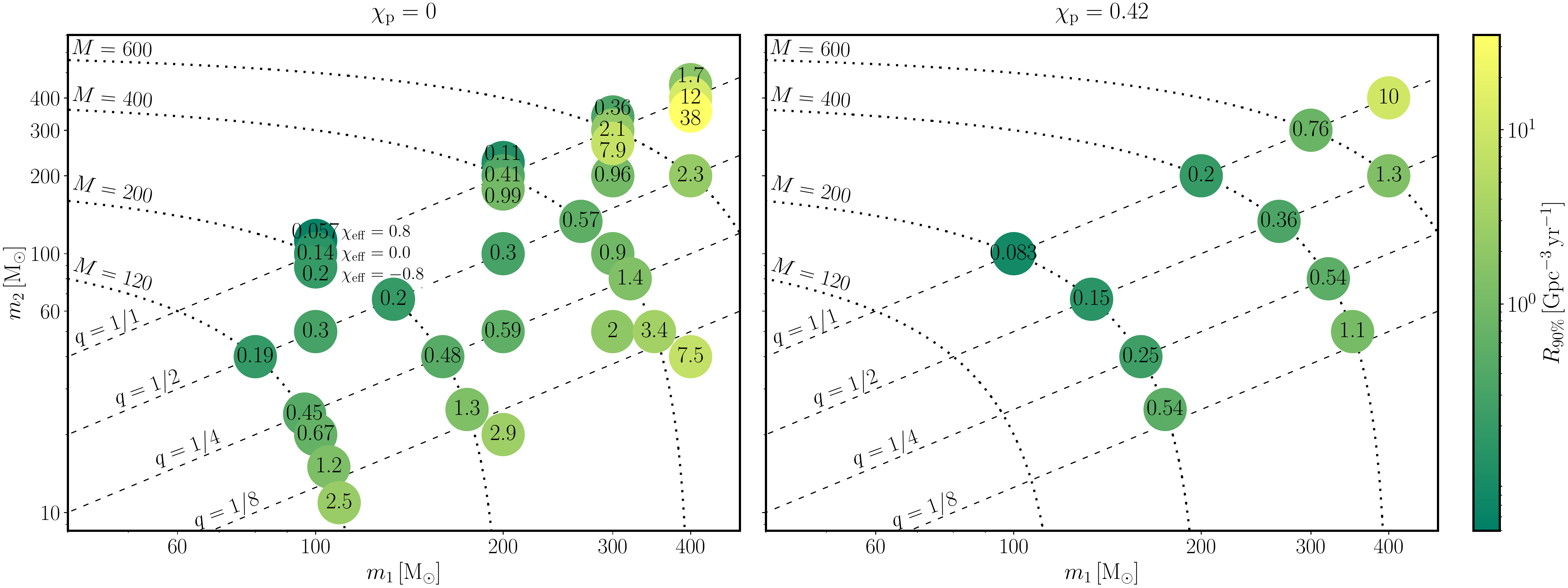}
    \caption{The 90\% upper limit on merger rate density $R_{90\%}$ in Gpc$^{-3}\,$yr$^{-1}$ for the targeted 
    \ac{IMBH} binary sources in the $m_1$-$m_2$ plane.  The values are rounded where necessary for display. Left panel for $\chi_\mathrm{p}=0$ and right panel for 
    $\chi_\mathrm{p}=0.42$. Each circle corresponds to one class of \ac{IMBH} binaries in the source frame. The 
    $\chi_\textrm{eff}$ values of injection sets are labelled and shown as displaced circles.}
    \label{Fig:Rate}
\end{figure*}

Here we calculate limits on the merger rate for points in the binary component 
mass and spin parameter space described in Table~\ref{Table:Injections},
using the loudest-event method~\citep{Biswas:2007ni,Abbott:2016ymx}. 

To derive the upper limit on merger rate for a given point in source parameter space, we consider
the sensitive volume-time, $\langle VT\rangle_{\mathrm{sen}}$, of our combined search to such sources at a p-value threshold of 0.251, which is determined by %the combined p-value of 
\valentinesdaynoise, the most significant event due to noise in the combined search results.  For mergers with 
given intrinsic parameters %defined by a point in mass-spin space, 
the expected number of detected signals $N$ is related to the merger rate $R$ and to the 
sensitive volume-time as $\langle N \rangle = R \langle VT\rangle_{\mathrm{sen}}$.
For each source point, we estimate $\langle VT\rangle_{\mathrm{sen}}$, as a fraction of the total volume-time
out to its maximum injection redshift $z_\mathrm{max}$, by counting injected signals that are 
detected with a combined p-value below the threshold and dividing by the total number of injections 
generated. 

Then, taking a uniform prior on $R$ and using the Poisson probability of zero detected signals 
as a likelihood, we obtain the $90\%$ credible upper limit $R_{90\%}=2.3/\langle VT \rangle_{\mathrm{sen}}$. 
The only significant IMBH binary signal in the combined search results is GW190521. However,
there is only one mass-spin (marked with \dag\ in Table~\ref{Table:Injections}) point which is 
consistent with both its component mass and spin 
$\chi_\mathrm{eff}$--$\chi_\mathrm{p}$ 90\% credible regions. Therefore, for that source point, we 
conservatively use the Poisson probability of having one IMBH binary detection and thus take
$R_{90\%}=3.9/\langle VT \rangle_{\mathrm{sen}}$.

Injections with component masses (60+60) \Msun were performed: however, since this parameter
point is within the stellar-mass \ac{BBH} distribution characterized in~\citet{o3apop}, to which several 
heavy BBH systems detected in O3a may contribute, we do not quote an upper rate limit.  We do, however, state search sensitivity for such systems in our data release products.  

Table~\ref{Table:Injections} summarises the sensitive volume-time and upper limit on the merger rate 
for our chosen set of injections. 
For simulated non-spinning sources, the sensitive volume-time decreases with an increase in total mass but increases with increasing mass ratio $q$. There are multiple reasons for these trends.  First, for a fixed mass ratio, the duration of a signal within the detector bandwidth decreases
with increased total mass, even though its overall intrinsic luminosity increases.  This is evident 
if one compares the sensitive volume-time obtained for $(80+40)$\,\Msun, $(100+50)$\,\Msun\, and $(133+67)$\,\Msun \,systems.
 Second, the amplitude of a source decreases with a decrease in the mass ratio for a fixed total mass. Hence the sensitivity drops with a decrease in mass ratio.  Last, a decrement in mass ratio also increases the contribution coming from sub-dominant emission multipoles. This significantly affects the GstLAL and PyCBC searches that filter using dominant multipole templates only.

Concerning the dependence on spins, for more positive (negative) values of the effective inspiral spin of a
system, keeping the source frame component masses fixed, the duration of the merger signal within the detector
bandwidth increases (decreases) as compared to a non-spinning counterpart. Hence the sensitivity improves 
(degrades) for systems with positive (negative) effective total spin~\citep{Salemi:2019ovz,Tiwari:2018qch}. 
All precessing systems used in this analysis have positive $\chi_\mathrm{eff}$: hence, the combined
search can observe them to a greater distance compared to their non-spinning counterpart.

Figure \ref{Fig:VTsens} shows this trend visually. The panels show the sensitive volume-time for 
non-precessing and precessing simulated sources, respectively. Each circle corresponds to one class of 
IMBH binaries in the source frame. The \ac{IMBH} binaries with aligned and anti-aligned BH
spins, $\chi_{1,2}$ are labeled and shown as displaced circles. In general, we find an 
increase in the sensitive volume-time of the combined search compared to results in~\citet{Salemi:2019ovz}. 
This increase is due to an overall increase in the analysis time, detector sensitivity, and the 
contributing searches' sensitivity. 

Figure \ref{Fig:Rate} shows the 90\% upper limit on merger rate, $R_{90\%}$, in \si{Gpc^{-3}\,yr^{-1}}
for the targeted 43 \ac{IMBH} binary sources in the $m_1$-$m_2$ plane. As before the left panel shows the 
result for non-precessing simulated sources whereas the right panel shows the same for precessing simulated 
sources. We set our most stringent upper limit 0.06 \si{Gpc^{-3}\,yr^{-1}} for equal-mass IMBH binaries 
with total mass $200\,\Msun$ and spin $\chi_{1,2}=0.8$ which is {$\sim 3.5$} times more stringent than the previous study \citep{Salemi:2019ovz}. 

\subsection{Updated GW190521 merger rate estimate}

We re-estimate the merger rate of a GW190521-like population. As in~\citet{TheLIGOScientific:2017qsa,
Abbott:2016ymx,GW190521Adiscovery, GW190521Aastro}, we consider a simulated signal to be detected if
it is recovered with an \ac{FAR} less than $100\,\mathrm{yr}^{-1}$. 
This corresponds to a combined p-value threshold of 0.009. We considered the maximum observed time 
($T_a = 0.874\,$yr) across the three pipelines as the analysis time of the combined search.
The population is generated by drawing the intrinsic parameters from the posterior distribution 
inferred using the \texttt{NRSur7dq4} waveform model~\citep{GW190521Aastro} and then distributing 
them isotropically over binary orientation parameters, sky position, and uniformly over comoving 
volume-time up to max redshift $z=1.5$. 

The sensitive volume-time of the combined search to GW190521-like mergers over the entire O3 data is 14.35\,\si{Gpc^3\,yr}. 
As in~\citet{GW190521Aastro} we take a Jeffreys prior proportional to $R^{-1/2}$, where $R$ is the astrophysical 
merger rate, and given the count of 1 detection above the threshold, obtain an estimate of 
$0.08^{+0.19}_{-0.07}$\,\si{Gpc^{-3}yr^{-1}} 
which is more constraining than the estimate given in~\citet{GW190521Aastro}, consistent with the higher observing 
time and increased sensitivity of the searches.  

%% file: conclusion.tex
\section{Discussion \& Conclusions }
\label{sec:conclusion}

The Advanced LIGO and Advanced Virgo detector network concluded their year-long third observing run
in March 2020. The first part of the run witnessed the first confident IMBH binary (GW190521; \citet{GW190521Adiscovery}) with the primary BH inferred to be in the PISN mass gap and remnant in the IMBH mass range.
In this work, we present the IMBH binary search carried out on the entire data of the third observing run. We use three \ac{GW} search algorithms for the same; the two template-based searches GstLAL and PyCBC, and one model waveform independent cWB search. The template-based matched filter searches use the dedicated template bank designed for the massive black hole binary coalescences, while the model waveform independent cWB uses a single detection approach for the entire range of BBHs masses including IMBH binaries. We rank the events based on the combined significance computed combining the three searches. The search shows that GW190521 is still the most significant IMBH binary event. Besides that, we do not find any other significant event in the combined search. We update the significance of already published O3a events with this combined search. We report the discussion on the candidate events, including the marginal events.

Amongst the remaining events, \januaryevent{} shows marginal significance in the offline cWB search. As the detector characterization study does not conclusively demonstrate the origin of this event to be terrestrial, we carry out a follow-up investigation on this event in Appendix-\ref{sec:200114}. This includes event reconstruction and residual analysis. While the pre-merger dynamics of \januaryevent{} are barely accessible, we analyse \januaryevent{} under the quasi-circular BBH hypothesis. Unlike the case of GW190521, here we find strongly inconsistent results across different waveform approximants.

But the model independent event reconstructions are consistent with each other. Hence, either the event is not consistent with the available quasi-circular binary black hole waveforms or its origin is non-astrophysical in nature. 
%\rough{We cannot confidently conclude this event corresponds to an astrophysical signal and do not consider it a detection.}

We update the merger rate density on a suite of numerical relativity signals of IMBH binary systems with generic BH spins using the numerical relativity waveforms provided by SXS, RIT, and GeorgiaTech catalogs. We compute the merger rate limit with the method of loudest confirmed noise trigger. The most stringent revised 90\% merger rate upper limit is placed on equal mass IMBH binary with an aligned spin of 0.8 to be {$0.056$ \si{Gpc^{-3}yr^{-1}}}. The improvement is a factor of {$\sim 3.5$}  over the earlier results using O1-O2 data. With the year-long O3 data and improved sensitivity of the combined search, we further revise the astrophysical merger rate estimates for binary systems comparable to GW190521 to {$0.08^{+0.19}_{-0.07}$ \si{Gpc^{-3}yr^{-1}}}. This is an improvement of factor 1.6 over the earlier result in ~\citet{GW190521Aastro}.  

We emphasize here that the IMBH binaries pose an extreme challenge to interpret. All current GW observations have been so far interpreted within the canonical scenario of an inspiraling quasi-circular BBH. While this is a safe assumption when the (pre-merger) inspiral process is clearly visible in the band, the low frequency of IMBH binary signals makes such putative inspiral barely visible. This leaves the pre-merger dynamics and even the very nature of the colliding objects open to further interpretation, making some conclusions obtained through canonical analyses less robust \citep{bustillo2020confusing}.
For instance, alternative scenarios have been proposed that GW190521 is consistent with an eccentric binary merger \citep{2020ApJ...903L...5R, Gayathri:2020coq}.
Moreover, \citet{PhysRevLett.126.081101} has
shown that GW190521 is consistent with the merger of exotic objects which has dramatically different astrophysical conclusions. The primordial BH scenario is explored by \cite{De_Luca_2021, clesse2020gw190425}. Alternative scenarios for forming BHs in the mass gap include gas accretion onto stellar mass BHs in dense molecular clouds, or in primordial dense clusters \citep{2019A&A...632L...8R, 2020ApJ...903L..21S, Rice_2021}.

With continuous improvement in advanced detectors especially in the low frequency region, we expect to probe more inspiral cycles of the high massive BBH systems as well as probe higher-order modes. Compounded with the improvement in the templated searches incorporating complex dynamics will improve the detectability of IMBH binaries with the matched filter based searches. Improved veto methods to distinguish between the short duration noisy transients with complex morphology from an IMBH signal is a useful step forward to detect more IMBH binaries.

The detection of massive BHs in the GW window has provided crucial observations input for the stellar evolutionary models. While we use the conservative limit of the lower edge of the mass gap in the BH population, it is highly uncertain, it might be as low as $\sim{}40$ M$_\odot$ or above $\sim{}70$ M$_\odot$, depending on uncertainties about the nuclear reaction rates \citep[e.g.,][]{2019ApJ...887...53F,2020ApJ...902L..36F,2021MNRAS.501.4514C}, the collapse of the residual stellar envelope \citep[e.g.,][]{2020ApJ...888...76M,2021MNRAS.501.4514C}, the impact of stellar rotation \citep[e.g.,][]{2020ApJ...888...76M,2020A&A...640L..18M, woosley2021pairinstability}, the result of stellar mergers \citep[e.g.,][]{2019MNRAS.487.2947D,2020MNRAS.498..495D,2020ApJ...904L..13R}, the efficiency of accretion from companion stars \citep[e.g.,][]{2020ApJ...897..100V}, the model of convection \citep[e.g.,][]{2020MNRAS.493.4333R,2021MNRAS.502L..40F, tanikawa2020population} 
and the onset of dredge-up episodes
 \citep[e.g.,][]{2021MNRAS.501.4514C, tanikawa2020population, Umeda_2020}. Recently, \citet{2021MNRAS.501.4514C} 
propose that the mass gap might even disappear if a low rate for the $^{12}{\textrm C}(\alpha{},\gamma{})^{16}{\textrm O}$ reaction is assumed and if a mild envelope under-shooting is included in stellar evolution calculations. Theoretical and numerical models show that black holes with mass in the pair instability gap could be the result of hierarchical mergers of smaller black holes \citep{2002MNRAS.330..232C,2016ApJ...831..187A, 2017PhRvD..95l4046G,2017ApJ...840L..24F,2019PhRvD.100d3027R, kimball2020evidence, 2020ApJ...893...35D} or the outcome of stellar collisions in dense star clusters \citep{2019MNRAS.487.2947D,2020MNRAS.497.1043D,2020ApJ...903...45K,2020ApJ...904L..13R}. Hierarchical mergers appear to be particularly efficient in nuclear star clusters \citep{2019MNRAS.486.5008A,2020ApJ...902L..26F,2020PhRvD.102d3002B,2021arXiv210305016M} and in the dense gaseous disks of active galactic nuclei \citep{2012MNRAS.425..460M,2017ApJ...835..165B,2018ApJ...866...66M,2019PhRvL.123r1101Y,2020ApJ...899...26T,2021ApJ...908..194T}. The detection of more massive BH binaries in the advanced detector era will provide constraints on all the formation channels. In addition, future observations of IMBH binaries across the GW spectrum \citep{Sathyaprakash_2009} could strengthen the possible evolutionary link between stellar-mass BHs and supermassive \acp{BH} in the galactic centres in coming decades \citep{Merzcua:2017, 2017mbhe.confE..51K,King:2004ri}.

%% file: appendix.tex
\section{Other marginal candidate events}
\label{sec:otherEvents}

\begin{table}[htb!]
\centering
\begin{tabular}{l|l|c|c|c|c}
\hline
 Event & GPS Time (s) & cWB \ac{FAR} (yr$^{-1}$)  & PyCBC \ac{FAR} (yr$^{-1}$) &  GstLAL \ac{FAR} (yr$^{-1}$)  & $\Bar{p}$ \\
 \hline
 \septembernoise &  1253402832.9 &       - & $9.0\times 10^{+1}$ & $4.0\times 10^{-1}$ & $6.5\times 10^{-1}$ \\
 \christmasevent &  1261346253.8 &       - & $4.7\times 10^{-1}$ & $2.0\times 10^{+3}$ & $6.5\times 10^{-1}$ \\
 \decembertwentythirdnoise &  1261100537.6 &       - &                   - & $4.6\times 10^{-1}$ & $7.0\times 10^{-1}$ \\
\hline
\end{tabular}
\caption{\label{Table:extra-candidates} Other marginal candidate event list. We find 3 candidate events that 
passed a \ac{FAR} threshold of 0.5\,yr$^{-1}$ in at least one of the three dedicated searches, and 
additionally have a combined p-value less than 0.7.}
\end{table}

Table~\ref{Table:extra-candidates} summarises the other marginal candidates identified 
by our combined \ac{IMBH} search in O3 data that are not reported elsewhere in a catalog of compact binary coalescence events and that satisfy the 
criteria described in Sect.~\ref{sec:imbh_definition}.  These triggers were 
reported by at least one of the contributing searches with a \ac{FAR} below 
0.5\,yr$^{-1}$, and have a $\Bar{p} \leq 0.7$. 

The event \christmasevent~was first reported by PyCBC Live~\citep{DalCanton:2020vpm}, 
a low-latency matched-filter search with a \ac{FAR} of $0.4\,\mathrm{yr}^{-1}$. 
When the contributing searches conducted a dedicated offline analysis, the PyCBC-IMBH search identified it with a \ac{FAR} of 
$0.47\,\mathrm{yr}^{-1}$. However, the cWB and GstLAL-IMBH searches did not identify the 
event.  The transient did not pass cWB veto threshold as discussed in Sect.~\ref{sec:cwb}.
The event is, though, identified by the most general cWB search for GW bursts of 
short duration with a \ac{FAR} of $\sim 2\,\mathrm{yr}^{-1}$, which is still consistent with noise origin.
The model-agnostic BayesWave (BW) analysis \citep{Cornish:2014kda, BayesWaveMethodsIII} also identified the event with a \ac{FAR}
of $\sim 1\,\mathrm{yr}^{-1}$. The event morphologically resembles the Tomte class of glitches~\citep{Davis:2021ecd} which are common but of unknown origin.
Followup of LIGO Livingston data also showed the existence of multiple comparable
glitches within $100\,$s of the event time. 

The remaining two events, \septembernoise~and \decembertwentythirdnoise, are likely caused by instrumental
noise. In the case of \septembernoise, fast light scattering noise extending up
to about 60 Hz is present in the LIGO Livingston data around the time of the event,
and there is a high-SNR glitch in the Virgo data. The time-frequency morphology
of the \decembertwentythirdnoise~signal in the LIGO Livingston data matches an instrumental glitch~\citep{Davis:2021ecd}.

%% file: s200114f.tex
\section{Followup studies of \januaryevent}
\label{sec:200114} 

The cWB offline search detects \januaryevent{} with a combined FAR of $0.058\,\mathrm{yr}^{-1}$. The network SNR with LHO-LLO network is 12.3 and the three detector network SNR is 14.5. Although we cannot exclude the terrestrial origin of \januaryevent{},   we did perform several follow-up studies on this candidate which we summarise here. The studies include event reconstruction by BayesWave (BW) and cWB, parameter estimation (PE) with models of black hole binary merger including effects of orbital precession and higher-order multipole emission, reconstruction of PE sample waveforms and comparison with the event reconstruction with cWB, and residual analysis with BW. 

\subsection{Investigation of instrumental noise }\label{sec:S200114fnoise}
As mentioned in Sect.~\ref{sec:SearchResults}, an instrumental noise transient at the Hanford observatory coincides with \januaryevent{}.
The noise originates from a fan on a laser controller located on top of a squeezed light optics enclosure.
At the time of the event, an accelerometer detected a second-long frequency dip in the 76\,Hz fan motion.
Such vibrational transients can weakly couple to the strain data through the squeezing system.
We perform a follow-up investigation using the methods described in~\citet{Nguyen:2021ybi} to acquire
accurate estimates of the vibrational coupling between the table accelerometer and the strain channel.
We estimate the expected noise in the strain channel at the fundamental frequency to be over an order
of magnitude below background levels, so the fan is highly unlikely to account for the event candidate;
however, the estimated noise at the first harmonic (152\,Hz) is about a factor of two below background
and could potentially impact parameter estimation.

\subsection{Event reconstruction by model independent analysis }
\begin{figure*}
	\hspace*{+0.1cm}
	\centering
	\includegraphics[scale=0.3]{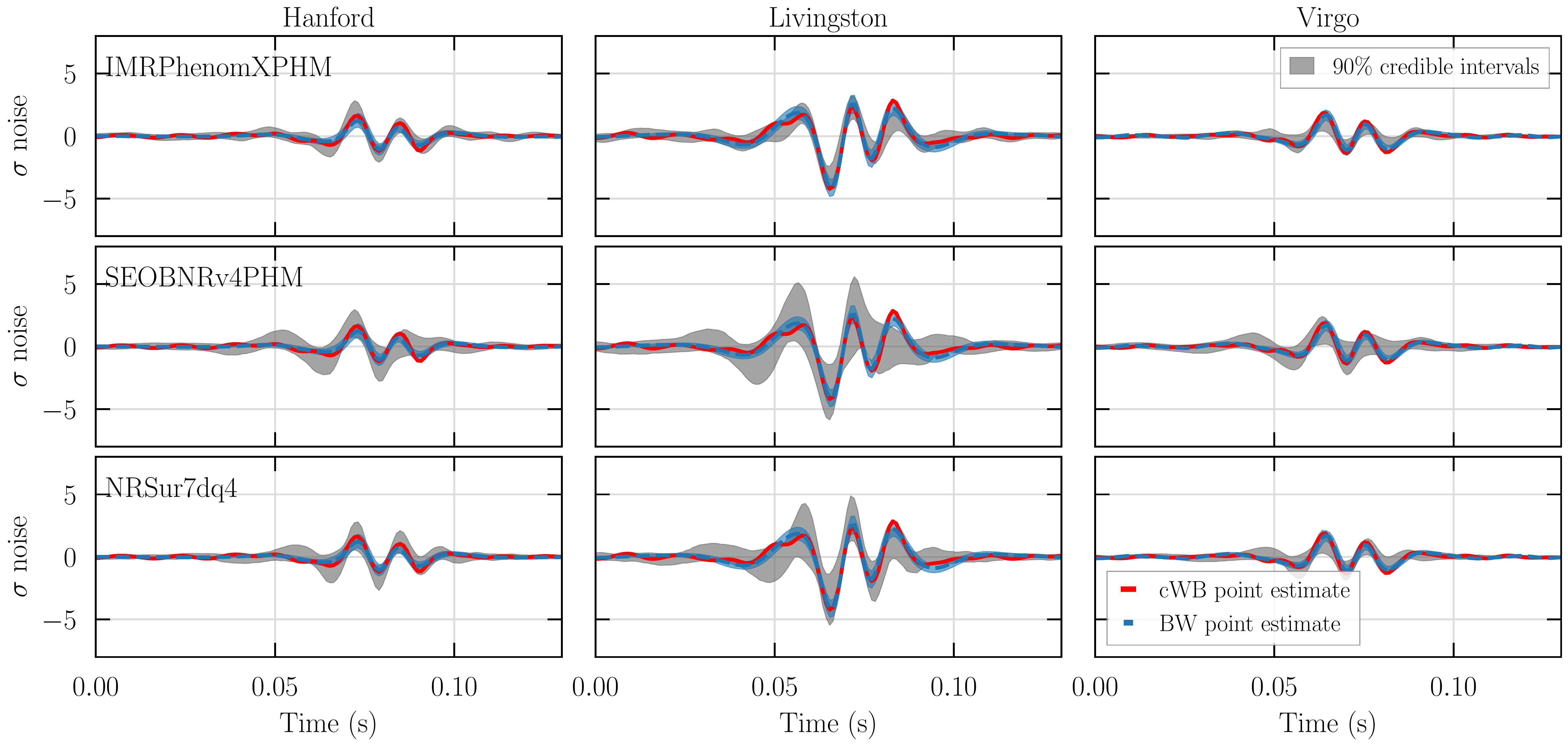}
	\caption{ The consistency of the waveform reconstruction by cWB with three waveform models in the time domain: IMRPhenomXPHM (upper panel), SEOBNRv4PHM (middle panel), and NRSur7dq4 (lower panel).  The coloured solid red and dashed blue curves correspond to the whitened reconstructed event by cWB and Bayeswave respectively.  
    The blue shaded region corresponds to the 90\% credible region from the event construction by BW. 
	The grey shaded belts are reconstructed waveforms by cWB for the 90\% credible interval corresponding to the PE runs.
	\label{Fig:rec}}
\end{figure*}

We reconstruct the signal using two model-independent analyses, namely cWB (used in the searches) and BW. The BW algorithm constructs the signal as a linear combination of sine-Gaussian wavelets and does not use any astrophysical model. The cWB reconstructs the multi-detector maximum likelihood signal by using the inverse wavelet transformation with selected pixels.  In Fig. \ref{Fig:rec}, the red coloured solid and dotted blue curves correspond to the whitened reconstructed signal from cWB and BW respectively. The cWB event reconstruction is within the 90\% credible region of the event reconstruction by BW (blue shaded region) for all three detectors. The BW SNR is 4, 14, and 5 in LHO, LLO and Virgo respectively, while cWB SNR is 5, 12, and 6 obtained from the reconstructed event. 

\subsection{PE analysis}
Here, we investigate the possibility that \januaryevent{}   may be described by the merger of a quasi-circular BBH system. We thus carry out parameter estimation with up-to-date waveform models including effects of precession and higher-order multipole moments. 
Specifically, we use three quasi-circular BBH waveform models $h(t;\theta)$: i) the numerical relativity surrogate model \texttt{NRSur7dq4} \citep{Varma:2019csw}; ii) the effective-one-body model \texttt{SEOBNRv4PHM} \citep{Ossokine:2020kjp, Babak:2016tgq} and iii) the phenomenological model \texttt{IMRPhenomXPHM} \citep{Pratten:2020ceb}. We perform the analysis on 8\,s of data centred around \januaryevent{}.  All analyses were performed on C01 60Hz subtracted data with a lower cutoff frequency of 10 Hz and reference frequency of 11 Hz. For the \texttt{IMRPhenomXPHM} analysis, we use the nested sampling algorithm as implemented in LALInference \citep{Veitch:2014wba}, while for \texttt{SEOBNRv4PHM} and \texttt{NRSur7dq4} analysis instead, we use the RIFT \citep{Lange:2017wki} analysis tool. 
Both algorithms are designed to compute the joint 15-dimensional posterior distribution $p(\theta|D,H)$ as well as the Bayes factor ($B_{\rm SN}$), Eq.~(\ref{eq:bayes_factor}), with prior distributions as described in the main text, Sect.~\ref{sec:imbh_definition}. The $\log_{10} B_{\rm SN}$ \footnote{The uncertainties on the individual $\log_{10} B_{\rm SN}$ are $\sim 1$.}
is tabulated in Table \ref{tab:S200114f-summary} for all the three runs. The values of  $\log_{10} B_{\rm SN}$ indicate a preference for the hypothesis H that a signal is present over the alternative of only Gaussian noise. 

These results do not, though, address the possibility that excess power in one or more detectors may be due to an instrumental artifact (glitch). As a diagnostic we therefore perform a \emph{coherence test} \citep{PhysRevD.78.022001},  
using the \texttt{IMRPhenomXPHM} waveform model. The coherence test computes the Bayes factor for the coherent signal hypothesis against the hypothesis of an incoherent signal in the network of detectors. It can be thus interpreted loosely as an indicator of the presence of accidentally coincident noise artefacts that could mimic an astrophysical signal. The resulting $\log_{10}$ Bayes factor for coherent vs.\ incoherent signal 
0.2, providing little to no evidence in support of the coherent signal hypothesis. Such small evidence is easily understood by looking at the $\log_{10}$ Bayes factors computed from analyses of each individual detector's data: both a Hanford-only as well as Virgo only analysis recovers a $\log_{10}$ Bayes factor for the signal vs.\ Gaussian noise hypothesis of 0.2. 
As a consequence, the posterior distributions from the Hanford- and Virgo-only analyses are largely uninformative. On the other hand, a Livingston-only analysis finds a $\log_{10}$ Bayes factor of 25. Hence, from the parameter estimation point of view, \januaryevent{} is essentially a single detector event. 
Returning to the results under the hypothesis $H$ of a quasi-circular merger signal plus Gaussian noise, we summarise the resulting median and symmetric 90\% credible regions for a few astrophysically relevant parameters from each of the models in Table~\ref{tab:S200114f-summary}.
\renewcommand{\arraystretch}{1.5}
\begin{table*}[tbh]
\centering
\begin{tabular}{c|c|c|c|c|c|c|c}
\hline
Waveform Model &              $m_1~(\Msun)$ &             $m_2~(\Msun)$ & $\chi_{\textrm{eff}}$ &  $\chi_{\textrm{p}}$ &                    $D_L~(\textrm{Mpc})$ &    $\theta_{\textrm{JN}}$ &  $\log_{10}B_{\textrm SN}$ \\
\hline
 \texttt{IMRPhenomXPHM} &  $197.2^{+22.6}_{-31.2}$ &  $62.1^{+11.1}_{-16.6}$ &  $-0.3^{+0.1}_{-0.2}$ &  $0.6^{+0.1}_{-0.2}$ &    $593.4^{+276.8}_{-146.7}$ &  $0.9 ^{+ 0.8 }_{ -0.3 }$ &                               24.9 \\
   \texttt{SEOBNRv4PHM} &  $179.6^{+17.1}_{-60.4}$ &     $26.7^{+7.3}_{-6.4}$ &  $-0.7^{+0.2}_{-0.1}$ &  $0.2^{+0.3}_{-0.2}$ &     $475.6^{+199.1}_{-153.0}$ &   $1.6^{+ 1.2 }_{ -1.1 }$ &                               26.2 \\
     %\texttt{NRSur7dq4} &   $77.2^{+33.4}_{-17.4}$ &  $41.7^{17.3}_{-17.5}$ &  $-0.5^{+0.3}_{-0.2}$ &  $0.5^{+0.3}_{-0.3}$ &  $1633.6^{+1099.8}_{-947.7}$ &  $2.0 ^{+ 0.8 }_{ -1.6 }$ &                               22.9 \\
\texttt{NRSur7dq4}& $ 75.0 ^{+ 32.9 }_{ -18.8 }$ & $ 42.5 ^{+ 16.4 }_{ -18.0 }$ & $ -0.5 ^{+ 0.3 }_{ -0.2 }$ & $ 0.5 ^{+ 0.3 }_{ -0.3 }$ & $ 1797.0 ^{+ 1601.0 }_{ -1027.0 }$ & $ 2.0 ^{+ 0.9 }_{ -1.6 }$ & 22.3\\
\hline
\end{tabular}
\caption{Summary of median and 90\% credible intervals of \januaryevent{} for different waveform models. The columns show the waveform model used for parameter estimation, the source frame component masses $m_i$, effective spin parameters $\chi_{\textrm{eff}}$ and $\chi_{\textrm{p}}$, luminosity distance $D_L$, the angle between the total angular momentum and the direction of propagation of the gravitational wave signal $\theta_{\textrm{JN}}$ and the $\log_{10}\textrm{Bayes~Factor}$ between the signal and Gaussian Noise given the model.}
\label{tab:S200114f-summary}
\end{table*}

\begin{figure}
    \hspace*{-2cm}
    \centering
    \includegraphics[scale=0.3]{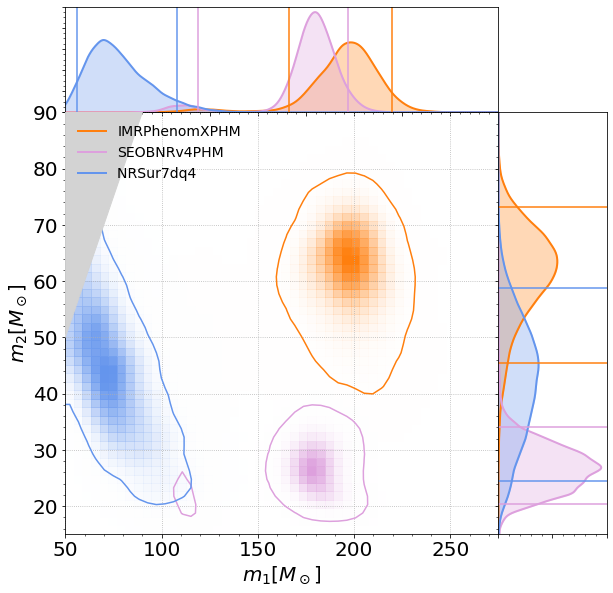}
     \includegraphics[scale=0.3]{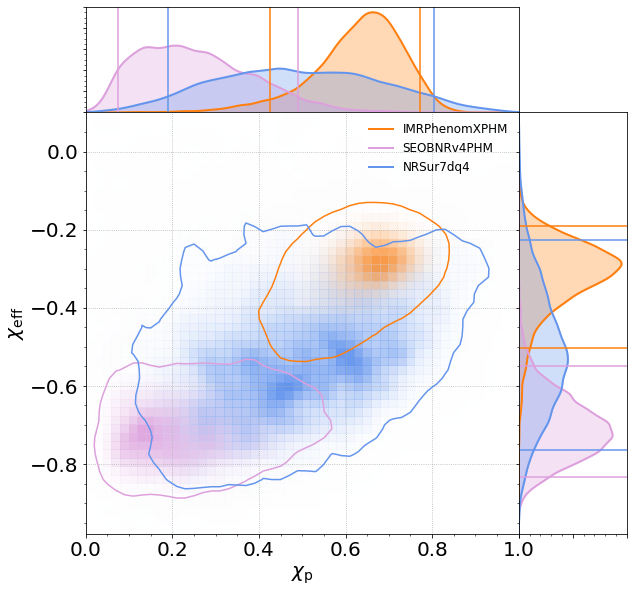}
    \caption{Posterior distributions: (Left) Source masses distribution, and (Right) the effective spin and effective in-plane spin distribution of \januaryevent{} for different waveform models. The 90$\%$ credible regions are indicated by the solid contour in the joint distribution and by solid vertical and horizontal lines in the marginalized distributions.}
    \label{Fig:source params}
\end{figure}

Fig.~\ref{Fig:source params} shows the joint posterior distribution for the component masses $m_1$ and $m_2$ of the source according to each waveform model. The three models infer BH masses that are largely inconsistent. In particular, the inferred values -- median and 90\% credible intervals -- show little overlap, see Table~\ref{tab:S200114f-summary}. Moreover, the result from \texttt{SEOBNRv4PHM} shows a hint of bimodality in the mass posterior distributions.

The posterior distributions for the spin parameters, Fig.~\ref{Fig:source params}, tell a similar story. If we compare the joint posterior distributions for the effective spin parameter $\chi_{\rm eff}$ along the direction of the orbital angular momentum and the in-plane effective spin parameter $\chi_{\rm p}$, we find that \texttt{IMRPhenomXPHM} and \texttt{SEOBNRv4PHM} probability distributions that are disjoint at the 90\% credible level. \texttt{NRSur7dq4} instead recovers a posterior distribution that is much broader and encompasses both the posterior from \texttt{IMRPhenomXPHM} and \texttt{SEOBNRv4PHM}. With reference to Table~\ref{tab:S200114f-summary}, all the three results indicate a preference towards the system being precessing and with their spin vectors anti-aligned compared to the orbital angular momentum. Spin vectors anti-aligned with the orbital angular momentum have the effect of accelerating the dynamical evolution of the system towards coalescence, resulting in shorter GW signals for a given chirp mass.

In summary, a follow-up investigation of the properties of \januaryevent{} interpreted as a possible quasi-circular binary merger shows considerable inconsistencies between results obtained by different waveform models. This is exemplified by the different posterior distributions for the BH masses as well as for their spins. Together with the lack of coherence among different detectors, our analysis indicates that, while we cannot exclude that \januaryevent{} has an astrophysical origin, there is no consistent support for its interpretation as a quasi-circular binary merger. 

\subsection{Residual analysis}
We evaluate the consistency of parameter estimation using the \texttt{IMRPhenomXPHM} waveform via two further analyses with BW. We obtain a match between the BW event reconstruction waveform with the maximum likelihood waveform from the \texttt{IMRPhenomXPHM} PE study as  0.871 which is consistent with expectations \citep{Ghonge:2020suv}.
Further, we perform a signal residual test by subtracting the maximum likelihood \texttt{IMRPhenomXPHM} waveform from the data and analyze the resulting residual with BW: the SNR obtained from the subtracted data is 7.4 for the LLO-LHO-Virgo network.
In parallel, we estimate the distribution of SNR expected in noise from different time segments of O3 data.  We estimate a p-value of {0.31} by comparing the distribution of noise \acp{SNR} with the actual event residual value. This study does not show significant evidence for excess noise.  

\subsection{Comparison between event reconstruction and injection recovery with PE samples}

Here, we compare the cWB reconstructed waveforms of the event against the waveforms estimated by the PE analysis~\citep{Szczepanczyk:2020osv,Gayathri:2020coq,PhysRevD.100.042003}.
We inject the waveforms corresponding to the PE samples into O3 data, estimate the reconstructed waveform for each of these samples using cWB and compare it with the reconstructed event by cWB. In Figure~\ref{Fig:rec}, the grey shaded belts are 90\% confidence intervals obtained with the cWB reconstruction of PE samples for each waveform. We observe that the cWB (solid red) and BW event (dashed blue) reconstructions are largely within this grey shaded belt.  Further, we note that the time-domain reconstruction with the \texttt{SEOBNRv4PHM} and \texttt{NRSur7dq4} samples have broader error belts as compared to the reconstruction with the \texttt{IMRPhenomXPHM} samples. This is possibly due to broad posteriors and errors in cWB reconstruction. 

To quantify the consistency between the cWB reconstruction and PE waveforms, we first compute the {\it null distribution} as the overlap between a injected waveform from the posterior distribution and its reconstructed waveform from the cWB.  The source distribution is the distribution of the overlap between injected waveform from the posterior sample and the cWB event reconstruction of \januaryevent{}. The spread in the null distribution owes to the cWB reconstruction, noise fluctuation and the posterior distribution. The spread in source distribution shows disagreement between the cWB event reconstruction and injected waveform from the posterior distribution. If the PE samples accurately describe any event, we expect a significant intersection between these two distributions. The p-value of the null distribution is the fraction of samples in the null distribution with the overlap below the overlap value of the maximum likelihood waveform with the reconstructed event. 
The p-values for \januaryevent{} are {0.01\%}, 0.4\% and {48\%} corresponding to \texttt{SEOBNRv4PHM}, \texttt{NRSur7dq4} and \texttt{IMRPhenomXPHM} posterior samples, with source overlaps for the maximum likelihood waveforms of {0.5}, {0.68} and {0.86} respectively. 
Thus, the low overlaps and p-values for the \texttt{SEOBNRv4PHM} and \texttt{NRSur7dq4} waveform models indicate that these models are inconsistent with the cWB reconstruction.

In a separate study, we inject \texttt{IMRPhenomXPHM} PE samples and recover the simulated events with cWB using the LHO-LLO and three detector networks.
We observe 25\% and 34\% injection recovery in LHO-LLO and LHO-LLO-Virgo configuration respectively.
The three detector network is recovering more events compared to the LHO-LLO network since LHO-LLO has blind/null spots in the sky. If the LHO-LLO-Virgo network saw the same population of signals as LHO-LLO network, then we would not re-analyze the data with the LHO-LLO-Virgo network.
73\% of the recovered injections by the three detector configuration is not recovered by LHO-LLO. 
However, out of the 34\% of samples recovered by the LHO-LLO-Virgo analysis, only 8.65\% of these samples have the same or higher significance than the \januaryevent{} event.
When we look at the detector network sensitivity skymaps, we find that it is not surprising that this event is missed by the LHO-LLO network.
We conclude that the significance estimated for \januaryevent{} is unlikely to result from a quasi-circular BBH with binary parameters according to \texttt{IMRPhenomXPHM}. 

The detailed analyses of \januaryevent{} under the quasi-circular BBH merger hypothesis gives inconsistent results across different waveform approximants with precession and higher-order multipole moments. This along with the residual study indicates that there is no consistent interpretation of the signal with the available quasi-circular merger waveforms.  
However, the unmodeled event reconstructions are consistent with each other. Hence, either the event is not consistent with the available quasi-circular binary black hole waveforms or its origin is non-astrophysical in nature. We do not report any alternate scenario such as eccentric binary merger due to lack of availability of waveforms which include both eccentricity and orbital precession.

%% file: LSC-Virgo-KAGRA-Authors-Feb-2021.tex
\overfullrule 0pt % delete the nasty little black boxes for overfull box
\parskip0pt
\parindent0pt
\hyphenpenalty9999

%\centerline{\bf LSC, Virgo and KAGRA February 2021 author list---LIGO-M2100041}
%\centerline{2021-04-29. Plain \TeX.}
\section{The LIGO Scientific Collaboration, the Virgo Collaboration, and the KAGRA Collaboration}
R.~Abbott,$^{1}$  %rich.abbott
T.~D.~Abbott,$^{2}$  %thomas.abbott
F.~Acernese,$^{3,4}$  %fausto.acernese
K.~Ackley,$^{5}$  %kendall.ackley
C.~Adams,$^{6}$  %carl.adams
N.~Adhikari,$^{7}$  %naresh.adhikari
R.~X.~Adhikari,$^{1}$  %rana.adhikari
V.~B.~Adya,$^{8}$  %vaishali.adya
C.~Affeldt,$^{9,10}$  %christoph.affeldt
D.~Agarwal,$^{11}$
% ...
%\par\medskip
%\centerline{(The LIGO Scientific Collaboration, the Virgo Collaboration, and the KAGRA Collaboration)}
%\par\medskip
%\parindent 0pt
%{${}^{\ast}$Deceased, August 2020. }%
%\medskip

$^{1}$LIGO Laboratory, California Institute of Technology, Pasadena, CA 91125, USA % {CIT} {1}
 
$^{2}$Louisiana State University, Baton Rouge, LA 70803, USA % {LSU} {2}
 
$^{3}$Dipartimento di Farmacia, Universit\`a di Salerno, I-84084 Fisciano, Salerno, Italy % {DIFASA} {3}
 
$^{4}$INFN, Sezione di Napoli, Complesso Universitario di Monte S. Angelo, I-80126 Napoli, Italy % {INFNNA} {4}
 
$^{5}$OzGrav, School of Physics \& Astronomy, Monash University, Clayton 3800, Victoria, Australia % {MonashU} {5}
 
$^{6}$LIGO Livingston Observatory, Livingston, LA 70754, USA % {LLO} {6}
 
$^{7}$University of Wisconsin-Milwaukee, Milwaukee, WI 53201, USA % {UWM} {7}
 
$^{8}$OzGrav, Australian National University, Canberra, Australian Capital Territory 0200, Australia % {ANU} {8}
 
$^{9}$Max Planck Institute for Gravitational Physics (Albert Einstein Institute), D-30167 Hannover, Germany % {AEIHannover} {9}
 
$^{10}$Leibniz Universit\"at Hannover, D-30167 Hannover, Germany % {Leibniz} {10}
 
$^{11}$Inter-University Centre for Astronomy and Astrophysics, Pune 411007, India % {IUCAA} {11}
 
% ...

%\end

%% file: P2000488_v5.tex
This material is based upon work supported by NSF’s LIGO Laboratory which is a major facility
fully funded by the National Science Foundation.
The authors also gratefully acknowledge the support of
% the United States National Science Foundation (NSF) for the construction and operation of the
% LIGO Laboratory and Advanced LIGO as well as
the Science and Technology Facilities Council (STFC) of the
United Kingdom, the Max-Planck-Society (MPS), and the State of
Niedersachsen/Germany for support of the construction of Advanced LIGO 
and construction and operation of the GEO600 detector. 
Additional support for Advanced LIGO was provided by the Australian Research Council.
The authors gratefully acknowledge the Italian Istituto Nazionale di Fisica Nucleare (INFN),  
the French Centre National de la Recherche Scientifique (CNRS) and
the Netherlands Organization for Scientific Research, 
for the construction and operation of the Virgo detector
and the creation and support  of the EGO consortium. 
The authors also gratefully acknowledge research support from these agencies as well as by 
the Council of Scientific and Industrial Research of India, 
the Department of Science and Technology, India,
the Science \& Engineering Research Board (SERB), India,
the Ministry of Human Resource Development, India,
the Spanish Agencia Estatal de Investigaci\'on,
the Vicepresid\`encia i Conselleria d'Innovaci\'o, Recerca i Turisme and the Conselleria d'Educaci\'o i Universitat del Govern de les Illes Balears,
the Conselleria d'Innovaci\'o, Universitats, Ci\`encia i Societat Digital de la Generalitat Valenciana and
the CERCA Programme Generalitat de Catalunya, Spain,
the National Science Centre of Poland and the Foundation for Polish Science (FNP),
the Swiss National Science Foundation (SNSF),
the Russian Foundation for Basic Research, 
the Russian Science Foundation,
the European Commission,
the European Regional Development Funds (ERDF),
the Royal Society, 
the Scottish Funding Council, 
the Scottish Universities Physics Alliance, 
the Hungarian Scientific Research Fund (OTKA),
the French Lyon Institute of Origins (LIO),
the Belgian Fonds de la Recherche Scientifique (FRS-FNRS), 
Actions de Recherche Concertées (ARC) and
Fonds Wetenschappelijk Onderzoek – Vlaanderen (FWO), Belgium,
the Paris \^{I}le-de-France Region, 
the National Research, Development and Innovation Office Hungary (NKFIH), 
the National Research Foundation of Korea,
the Natural Science and Engineering Research Council Canada,
Canadian Foundation for Innovation (CFI),
the Brazilian Ministry of Science, Technology, and Innovations,
the International Center for Theoretical Physics South American Institute for Fundamental Research (ICTP-SAIFR), 
the Research Grants Council of Hong Kong,
the National Natural Science Foundation of China (NSFC),
the Leverhulme Trust, 
the Research Corporation, 
the Ministry of Science and Technology (MOST), Taiwan,
the United States Department of Energy,
and
the Kavli Foundation.
The authors gratefully acknowledge the support of the NSF, STFC, INFN and CNRS for provision of computational resources.

%{\bf For papers using O3b (and future) data, the following paragraph should be added for KAGRA.}\\
This work was supported by MEXT, JSPS Leading-edge Research Infrastructure Program, JSPS Grant-in-Aid for Specially Promoted Research 26000005, JSPS Grant-in-Aid for Scientific Research on Innovative Areas 2905: JP17H06358, JP17H06361 and JP17H06364, JSPS Core-to-Core Program A. Advanced Research Networks, JSPS Grant-in-Aid for Scientific Research (S) 17H06133, the joint research program of the Institute for Cosmic Ray Research, University of Tokyo, National Research Foundation (NRF) and Computing Infrastructure Project of KISTI-GSDC in Korea, Academia Sinica (AS), AS Grid Center (ASGC) and the Ministry of Science and Technology (MoST) in Taiwan under grants including AS-CDA-105-M06, Advanced Technology Center (ATC) of NAOJ, and Mechanical Engineering Center of KEK. 

%{\bf For certain collaboration papers, it may be appropriate to acknowledge specific analysis software.
%One template to consider is that used for the GW190425 discovery paper, for which the latex can be
%found here:}\\
%{\small https://git.ligo.org/publications/gw190425/gw190425-discovery/-/blob/master/gw190425-discovery.tex\#L259.}

%{\bf For collaboration papers released after March 2020, it may be appropriate to add the following special acknowledgement, BUT the journal may reject it (PRL rejected it). If one chooses, one can include such acknowledgements in arXiv/DCC versions only in that case:}\\
{\it We would like to thank all of the essential workers who put their health at risk during the COVID-19 pandemic, without whom we would not have been able to complete this work.}